\documentclass[fleqn,usenatbib]{mnras}

\usepackage[T1]{fontenc}
\usepackage{ae,aecompl}

\DeclareRobustCommand{\VAN}[3]{#2}
\let\VANthebibliography\thebibliography
\def\thebibliography{\DeclareRobustCommand{\VAN}[3]{##3}\VANthebibliography}

\usepackage{graphicx}	
\usepackage{amsmath}	
\usepackage{amssymb}	
\usepackage{color}
\usepackage{algorithm}
\usepackage{booktabs}
\usepackage[noend]{algpseudocode}
\usepackage{pifont}
\usepackage{soul}
\usepackage{float}
\usepackage{multibib}
\newcites{Appendix}{Appendix References}
\usepackage{natbib}
\usepackage[flushleft]{threeparttable}
\usepackage{mathrsfs}

\graphicspath{{Figures/}}

\newcommand{\vecB}[1]{\mathrm{{\bmath{\mathit{#1}}}}}
\newcommand{\Exp}[1]{\mathrm{\left\langle #1 \right\rangle}}

\renewcommand{\d}[1]{\ensuremath{\operatorname{d}\!{#1}}}

\DeclareMathOperator\M{\mathcal{M}}
\DeclareMathOperator\Ma{\mathcal{M}_{\text{A}}}
\DeclareMathOperator\Mao{\mathcal{M}_{\text{A0}}}
\newcommand{\MaO}[1]{\mathcal{M}^{#1}_{\text{A0}}}

\defcitealias{Brunt2010a}{BFP2010}
\defcitealias{Beattie2020}{BF2020}
\defcitealias{Kolmogorov1941}{K41}
\defcitealias{Menon2020}{MFK2020}
\defcitealias{Federrath2016c}{F16}

\title[Anisotropic magnetic field fluctuations]{Magnetic field fluctuations in anisotropic, supersonic turbulence}

\author[J. R. Beattie, C. Federrath \&  A. Seta]{
James R. Beattie$^{1}$\thanks{E-mail: james.beattie@anu.edu.au},
Christoph Federrath$^{1}$ and Amit Seta$^{1}$
\\
$^{1}$Research School of Astronomy and Astrophysics, Australian National University, Canberra, ACT 2611, Australia
}

\date{Accepted 2020 July 24. Received 2020 July 18; in original form 2020 May 28.}

\pubyear{2020}

\begin{document}
\label{firstpage}
\pagerange{\pageref{firstpage}--\pageref{lastpage}}
\maketitle

\begin{abstract}
The rich structure that we observe in molecular clouds is due to the interplay between strong magnetic fields and supersonic (turbulent) velocity fluctuations. The velocity fluctuations interact with the magnetic field, causing it too to fluctuate. Using numerical simulations, we explore the nature of such magnetic field fluctuations, $\delta\vecB{B}$, over a wide range of turbulent Mach numbers, $\M = 2 - 20$ (i.e., from weak to strong compressibility), and Alfv\'en Mach numbers, $\Mao = 0.1 - 100$ (i.e., from strong to weak magnetic mean fields, $B_0$). We derive a compressible quasi-static fluctuation model from the magnetohydrodynamical (MHD) equations and show that velocity gradients parallel to the mean magnetic field give rise to compressible modes in sub-Alfv\'enic flows, which prevents the flow from becoming two-dimensional, as is the case in incompressible MHD turbulence. We then generalise an analytical model for the magnitude of the magnetic fluctuations to include $\M$, and find $|\delta\vecB{B}| = \delta B = c_s\sqrt{\pi\rho_0}\M\Mao$, where $c_s$ is the sound speed and $\rho_0$ is the mean density of gas. This new relation fits well in the strong $B$-field regime. We go on to study the anisotropy between the perpendicular ($ B_{\perp}$) and parallel ($ B_{\parallel}$) fluctuations and the mean-normalised fluctuations, which we find follow universal scaling relations, invariant of $\M$. We provide a detailed analysis of the morphology for the $\delta  B_{\perp}$ and $\delta  B_{\parallel}$ probability density functions and find that eddies aligned with $B_0$ cause parallel fluctuations that reduce $B_{\parallel}$ in the most anisotropic simulations. We discuss broadly the implications of our fluctuation models for magnetised gases in the interstellar medium. 
\end{abstract}

\begin{keywords}
MHD -- turbulence -- ISM: kinematics and dynamics -- ISM: magnetic fields -- ISM: structure
\end{keywords}



\section{Introduction}\label{sec:intro}

    Magnetised plasmas that undergo random velocity fluctuations are ubiquitous in the Universe. For cool, molecular clouds, the birthplace of stars, the velocity fluctuations are supersonic \citep{Larson1981,Solomon1987,Padoan1997,Ossenkopf2002,Elmegreen2004,Heyer2004,MacLow2004,Krumholz2005,Ballesteros2007,RomanDuval2011,Federrath2013,Schneider2013,Chevance2020}. But random velocity fluctuations alone cannot explain all of the structure that we observe in these clouds. Indeed, strong magnetic fields facilitate and help set the structure of the neutral hydrogen clouds \citep{Planck2016a,Planck2016b,Hennebelle2019,Krumholz2019,Clark2019}. For example, magnetic fields create magnetosonic striations in star-forming clouds \citep{Tritsis2016,Tritsis2018,Tritsis2018b,Beattie2020} and in general, contribute to changing the density dispersion of the clouds through magnetic cushioning \citep{Molina2012,Mandal2020}. Magnetic fields polarise the density structures in molecular clouds and hydrodynamical shocks preferentially form across magnetic field lines \citep{Clark2015,Planck2016a,Planck2016b,Cox2016,Malinen2016,Soler2017,Juan2019,Beattie2020,Clark2019,Seifried2020}. Magnetic fields inhibit cloud fragmentation, in turn influencing the initial mass function for new-born stars in the modern star-forming era \citep{Price2008,Hennebelle2011,Federrath2012,Federrath2015,Krumholz2019} and the era of first stars \citep{Sharda2020}. They facilitate anisotropic cloud collapse through flux-freezing of the initial mass-to-flux ratio, which causes molecular clouds to preferentially collapse parallel to large scale magnetic fields \citep{Tritsis2015,Mocz2017,Mocz2018}. At smaller scales magnetic fields also facilitate the launching of protostellar jets and outflows \citep{Frank2014,Kuruwita2017,Gerrard2019,Krumholz2019,Kurwita2019}, and influence the accretion rate of newly forming stars \citep{Kurwita2020}. This means, if one wants to study the dynamics of star-forming molecular clouds (MCs), and indeed calculate the potential that a cloud has to form stars, one must understand the properties of magnetised, supersonic turbulence \citep{Elmegreen2004,MacLow2004,McKee2007,Hennebelle2012,Padoan2014,Federrath2012,Federrath2013b,Federrath2018}. 

\subsection{Magnetised Turbulence}

    Magnetohydrodynamical (MHD) turbulence comes in a number of different flavours, largely depending on the strength of the (kinetic) turbulence compared to the strength of the magnetic field and whether or not the flow is compressible \citep[for a modern review]{Beresnyak2019}. One can encode this information into the Alfv\'en Mach number,
    \begin{equation} \label{eq:Ma}
        \Ma = \frac{V}{V_A} = \sqrt{4\pi\rho}\frac{V}{B}
    \end{equation}
    where $V$, $V_A = B / \sqrt{4\pi\rho}$, $\rho$ and $B$ are the root-mean-squared (rms) velocity, Alfv\'en velocity, density and magnetic field strength, respectively. For $\Ma < 1$, $\rho V^2 \lesssim B^2$ the magnetic field contributes significantly to the dynamics of the flow through the Lorentz force. This is the sub-Alfv\'enic regime. For $\Ma > 1$, $\rho V^2 \gtrsim B^2$ the magnetic field plays a lesser role and the turbulent motions set, for example the statistics of the flow \citep{Padoan2011,Molina2012,Beattie2020}. This is called the super-Alfv\'enic regime, and the trans-Alfv\'enic regime, $\Ma \sim 1$, separates the two. In our study we explore a broad range of $\Ma$, from $0.1 - 100$, to include the two dynamically dissimilar regimes. 
    
    Turbulent magnetic fields can be separated into two components,
    \begin{align} \label{eq:fluctuate0}
        \vecB{B}(t) = \vecB{B}_0(t) + \delta\vecB{B}(t),
    \end{align}
    where $\vecB{B}$ is the total field, $\vecB{B}_0$ is the ordered component of the field, and $\delta\vecB{B}$ is the turbulent (or fluctuating) component of the field (sometimes called $\vecB{B}_{\text{turb}}$ or $B_{\text{t}}$ in some literature; \citealt{Pillai2015,Federrath2016c}). The ordered component of the magnetic field can extend over an entire molecular cloud, creating coherent magnetic field structures at the parsec scale that evolve on much longer dynamical times than the fluctuating component of the field \citep{Bertrang2014,Pillai2015,Federrath2016a,Hu2019}. We explore the case in our study where the ordered component does not change in time at all, and the fluctuating component evolves self-consistently with the MHD equations, in a statistically stationary state, hence, we can rewrite Equation~(\ref{eq:fluctuate0}) as
    \begin{align} \label{eq:fluctuate}
        \vecB{B}(t) = \vecB{B}_0 + \delta\vecB{B}(t),
    \end{align}
    where $\Exp{\vecB{B}(t)}_t = |\vecB{B}_0|$ which means $\Exp{\delta\vecB{B}(t)}_t = 0$. For this reason, we will refer to $\vecB{B}_0$ as the mean-field throughout this study. 

\subsection{The Alfv\'en Mach Number of MCs}

    Since the magnetic field has two distinct components one can talk about the mean-field or fluctuating Alfv\'en Mach number. In this study, we find the dependence of the magnetic field fluctuations, $\delta \vecB{B}$, on the Alfv\'en Mach number defined with respect to the mean-field. We define the mean-field Alfv\'en Mach number exactly as we defined the rms $\Ma$ in Equation~(\ref{eq:Ma}), but with mean-field components, $\Mao = V / V_{\text{A}0} = (\sqrt{4\pi\rho_0} V) / B_0$, where $B_0 = |\vecB{B}_0|$ and $V_{\text{A}0}$ is the velocity of Alfv\'en waves along the mean-field. We explore a wide parameter set that encompasses both the sub- and super-Alfv\'enic mean-field regime, covering the parameter space of observed molecular clouds. 
    
    For example, for the central molecular zone cloud, G0.253+0.016 studied in \citet{Federrath2016b}, they measure a mean-field of $B_0 = (2.07 \pm 0.95) \,\text{mG}$ \citep{Pillai2015}, a mean volume density of $\rho_0 = (6.2 \pm 3.3)\times 10^{-20} \, \text{g/cm$^3$}$ and a velocity dispersion of $V = (6.8 \pm 0.2) \, \text{km/s}$ (see \citealt{Federrath2016b} for details on assumptions and further references). Using these quantities and Equation~(\ref{eq:Ma}) one can calculate $V_{A0} = (24 \pm 17) \,\text{km/s}$ which means that $\Mao = 0.3 \pm 0.2$, placing it well into the sub-Alfv\'enic mean-field regime. Furthermore, using the direction of the velocity gradients to infer the magnetic field strength, \citet{Hu2019} measure the Alfv\'en Mach number for five star-forming clouds in the Gould Belt: Taurus ($\Mao = 1.19 \pm 0.02$), Perseus A ($1.22 \pm 0.05$), L 1551 ($0.73 \pm 0.13$), Serpens ($0.98 \pm 0.08$) and NGC 1333 ($0.82 \pm 0.24$). Hence, the trans to sub-Alfv\'enic magnetised turbulence regime is of great importance for understanding the environment that stars form in, which is the focus of this study.  
    
    Previous work by \citet[herein called \citetalias{Federrath2016c}]{Federrath2016c} explored the magnitude of magnetic field fluctuations across a broad range of $B_0$ and $\delta B$ values, but for a single $\M$ value. In our study we will expand upon some of the key results in \citetalias{Federrath2016c} and elucidate how the $\M$ number influences the fluctuating magnetic field. Of particular importance for our study is the \citetalias{Federrath2016c} analytical model for $\delta B$ in the strong-field regime, where $B_0 \gg \delta B$. By relating the turbulent magnetic energy density, $e_\text{m} \approx B_0 \delta B / (4\pi)$ and the turbulent kinetic energy $e_\text{k} = (\rho_0 V^2) / 2$, \citetalias{Federrath2016c} found that
    \begin{align} \label{eq:fed2016mod}
        \delta B = \MaO{2} B_0 / 2,
    \end{align}
    by assuming that all of the turbulent magnetic energy was fed from the kinetic energy in the plasma, i.e., $e_{\text{m}} = e_{\text{k}}$. We expand upon this study significantly by exploring the fluctuations perpendicular and parallel to $\vecB{B}_0$ separately in high-resolution MHD simulations, by showing the mean normalised fluctuations are independent of $\M$, and by deriving a new model directly from the induction equation in the compressible, ideal MHD model of plasmas. Hence our study contributes significantly to better understanding the nature of magnetic fluctuations in compressible, astrophysically relevant flows.
    
    The study is organised into the following sections. First, in \S\ref{sec:theory} we derive a compressible quasi-static model for the fluctuations of the magnetic field from the induction equation, which is relevant for understanding how the magnetic anisotropy influences the dynamics of molecular clouds with a strong mean magnetic field present. We use our model to show that velocity gradients parallel to the magnetic field give rise to compressible modes in the turbulence, and we provide an order of magnitude estimate for the fluctuations as a function of $\M$. In \S \ref{sec:Sims} we discuss the simulations that we use to explore the magnetic field fluctuations and test our models. In \S \ref{sec:CompressFlow} we qualitatively explore how velocity gradients parallel to the mean magnetic field create compressible modes in the velocity field and discuss why this is relevant to the analysis of astrophysical observations. In \S \ref{sec:1pointStats} we test our fluctuation model on the simulation data, and show it reduces to the \citetalias{Federrath2016c} analytical model for the strong-field fluctuations. We explore the anisotropy and mean-normalised fluctuations, which show universal scaling laws that do not depend upon $\M$. We then explore the morphology of the magnetic field probability density functions (PDFs) parallel and perpendicular to the mean-field. Finally in \S\ref{sec:conclusion} we summarise our key findings.

\section{Compressible Magnetic Fluctuation Model}\label{sec:theory}

\subsection{Model derivation}

    In this section we derive a model for the amplitude of the fluctuations directly from the induction equation in the ideal MHD framework. We use the model derived in this section to explain the magnetic field fluctuations in the remainder of the study. The ideal, isothermal MHD equations are
    \begin{align}
        \frac{\partial \rho}{\partial t} + \nabla\cdot(\rho \vecB{v}) &= 0 \label{eq:continuity}, \\
        \left( \frac{\partial}{\partial t} + \vecB{v}\cdot\nabla \right) \rho\vecB{v} &= \frac{(\vecB{B} \cdot \nabla)\vecB{B}}{4\pi} - \nabla \left(c_s^2 \rho + \frac{|\vecB{B}|^2}{8\pi}\right) + \rho \vecB{F},\label{eq:momentum} \\ 
        \frac{\partial \vecB{B}}{\partial t} &= \nabla \times (\vecB{v} \times \vecB{B}), \label{eq:induction} \\
        \nabla \cdot \vecB{B} &= 0, \label{eq:div0}
    \end{align}
    where $\vecB{v}$ is the fluid velocity, $\rho$ the density, $\vecB{B}$ the magnetic field, $c_s$ the sound speed and $\vecB{F}$, a stochastic driving function that gives rise to the turbulence (which could be from, for example, supernova shocks permeating through the interstellar medium, or internal to the MC, gravity, galactic-scale shocks or ambient pressure from the galactic environment; \citealt{Brunt2009,Elmegreen2009IAUS,Federrath2015a,Krumholz2016,Grisdale2017,Jin2017,Kortgen2017,Federrath2017IAUS,Colling2018,Schruba2019,Lu2020}). We consider a magnetic field of the form shown in Equation~(\ref{eq:fluctuate}), with constant mean component of the field in the $z$-direction,
    \begin{align}\label{eq:magField}
    \begin{pmatrix}
    B_x(t) \\ 
    B_y(t) \\ 
    B_z(t)
    \end{pmatrix} &= 
    \begin{pmatrix}
    0 \\ 
    0 \\ 
    B_0
    \end{pmatrix} +
    \begin{pmatrix}
    \delta B_x(t) \\ 
    \delta B_y(t) \\ 
    \delta B_z(t)
    \end{pmatrix}.
    \end{align}
    This gives rise to two natural length scales in the vector field: one perpendicular to $\vecB{B}_0$, and one parallel to $\vecB{B}_0$. This is because the mean magnetic field defines an axis of symmetry in the turbulent flow \citep{Cho2002}. We will denote the two scales with $\perp$ (perpendicular) and $\parallel$ (parallel) subscripts throughout this study, respectively. To develop some insight into the nature of the magnetic fluctuations we now explore the time evolution of Equation~(\ref{eq:magField}) by propagating it through the induction equation (Equation~\ref{eq:induction}). By construction, one can show that the induction equation only governs the time evolution of the fluctuating component of the field, since the mean-field is time-independent and, since the mean-field is only in the $z$ direction, the $\nabla \times (\vecB{v} \times \vecB{B})$ term can be significantly simplified (we show the full derivation of this in Appendix \ref{appendix:induction}), to show that
    \begin{align} \label{eq:induction2}
        \left(\frac{\partial }{\partial t} + \vecB{v}\cdot\nabla\right)\delta\vecB{B} = & \overbrace{B_0 \partial_{\parallel} \vecB{v}}^{\text{$v_z$-gradient}} + \overbrace{(\delta\vecB{B} \cdot \nabla)\vecB{v}}^{\text{advection}} \nonumber \\ 
        & - \underbrace{(\vecB{B}_0 + \delta\vecB{B})(\nabla \cdot \vecB{v})}_{\text{compression}},
    \end{align}
    which reveals that in the Lagrangian frame of the fluid, the magnetic field fluctuations are influenced by (1) the velocity gradients along the mean magnetic field (which we indicate with $\partial_{\parallel}$, i.e., they are anisotropic), (2) the advection of the velocity by the fluctuations, and (3) the compression of the velocity field, assuming $\vecB{B}_0$ is independent of time. Disregarding the compression term, $(\vecB{B}_0 + \delta\vecB{B})(\nabla \cdot \vecB{v})$, and as $\partial_t\delta B \rightarrow 0$, this form of the induction equation is known as the quasi-static approximation of the magnetic field perturbation, which
    is only applicable in the case of incompressible plasma, \citep[and references therein]{Zikanov1998,Verma2017}, but here we have generalised the equation for compressible MHD plasmas, relevant for astrophysical phenomena. 
    
    For sub-Alfv\'enic molecular clouds, such as G0.253+0.016, $B_0 \gg \delta B$ (in this case, $B_0 > \delta B$ by an order of magnitude; \citealt{Federrath2016b}), and the time derivative of $\delta B$ is very small, i.e., much smaller than the time scale of compression. Hence, by assuming $B_0 \gg \delta B$ and $(D/Dt)\delta B \approx 0$,\footnote{$\frac{D}{Dt} \equiv \frac{\partial}{\partial t} + \vecB{v} \cdot \nabla$ is the material derivative.}  Equation~(\ref{eq:induction2}) simply becomes
    \begin{align} \label{eq:induction2.2}
        \frac{\vecB{B}_0}{B_0} (\nabla \cdot \vecB{v}) = \partial_{\parallel} \vecB{v},  
    \end{align}
    and by taking the magnitudes of both sides,
    \begin{align}\label{eq:induction2.1}
        |\nabla \cdot \vecB{v}| = |\partial_{\parallel} \vecB{v}|.
    \end{align}
    Thus, it is primarily the velocity streams that run parallel with the mean magnetic field that give rise to compressive modes in the clouds\footnote{Note that these are in addition to the mixture of various modes caused by the turbulent driving \citep{Federrath2008,Federrath2010}.}. We show the full derivation of Equation~(\ref{eq:induction2.1}) in Appendix \ref{appendix:vGradient}. This is consistent with other studies, where hydrodynamical shocks appear perpendicular to the mean magnetic field, with the compression happening in the parallel direction to the mean magnetic field \citep{Mocz2018,Beattie2020,Seifried2020}. We will discuss this further in \S\ref{sec:CompressFlow}.
    
    Equation~(\ref{eq:induction2.1}) shows the key difference between incompressible and compressible MHD turbulence for the case when $B_0 \gg \delta B$. If, for example, $ \nabla \cdot \vecB{v} = 0$, the velocity gradient along the field disappears, and hence the magnetic field fluctuations cannot persist along the field. This is known as the two-dimensionalisation of the three-dimensional flow \citep{Alexakis2011,Verma2017}, which for supersonic turbulence, is forbidden to happen as shocks form and travel parallel to the field from the parallel velocity gradient. This means magnetic field fluctuations persist along the field, and the flow retains its three-dimensional nature. We will see this is indeed the case for parallel fluctuations in numerical simulations (demonstrated in \S\ref{sec:1pointStats}).

\subsection{An order of magnitude estimate for $\delta B$}

    We now consider the dimensionless form of Equation~(\ref{eq:induction2}), where length scales, $\hat{\vecB{\ell}}= \vecB{\ell} / L$, velocities $\hat{\vecB{v}} = \vecB{v} / V$, and magnetic fields, $\delta \hat{\vecB{B}} = \delta\vecB{B} / \delta B$, $\hat{\vecB{B}}_0 = \vecB{B}_0 / B_0$, are normalised by their magnitudes, and the time-scale in the derivative we chose to be $\hat{t} = t / T$, where $T = L / V_{A0}$, is the time-scale of fluctuations travelling along the mean magnetic field that spans the full system scale $L$. By grouping $\delta\vecB{B}$ and $\vecB{B}_0$ terms on the right hand side of Equation~(\ref{eq:induction2}) and by applying our non-dimensionalisation, we find,
    \begin{align} \label{eq:induction3}
        \frac{\delta B V_{A0}}{L}\frac{\partial \delta\vecB{\hat{B}}}{\partial \hat{t}} & + \frac{\delta B V}{L}( \vecB{\hat{v}}\cdot\hat{\nabla} )\delta\vecB{\hat{B}} = \nonumber \\ 
        \frac{B_0V}{L} \left[ \hat{\partial}_{\parallel} \vecB{\hat{v}} - \vecB{\hat{B}}_0 (\hat{\nabla} \cdot \vecB{\hat{v}}) \right] + & \frac{\delta BV}{L}\left[(\delta\vecB{\hat{B}}\cdot \hat{\nabla})\vecB{\hat{v}}
         - \delta\vecB{\hat{B}}(\hat{\nabla} \cdot \vecB{\hat{v}})\right]. 
    \end{align}
    If we assume that $\hat{\partial}_{\parallel} \vecB{\hat{v}} - \vecB{\hat{B}}_0 (\hat{\nabla} \cdot \vecB{\hat{v}})$ and $(\delta\vecB{\hat{B}}\cdot \hat{\nabla})\vecB{\hat{v}}  - \delta\vecB{\hat{B}}(\hat{\nabla} \cdot \vecB{\hat{v}})$ are $\mathcal{O}(1)$, which is true by construction, then we can make an order-of-magnitude estimate of the fluctuations,
    \begin{align} \label{eq:induction4}
        \frac{\delta B V_{A0}}{L} + \frac{\delta B V}{L} \sim  \frac{B_0V}{L} + \frac{\delta BV}{L}, \\
        V_{A0} + V \sim  \frac{B_0}{\delta B} V + V, \\
        \delta B \sim  B_0 \Mao. 
    \end{align}
    Since $B_0 = 2c_s\sqrt{\pi\rho_0}(\M / \Mao)$ by definition, we ultimately find
    \begin{align} \label{eq:beattieModel}
        \delta B =  2c_s\sqrt{\pi\rho_0} C \M,
    \end{align}
    where $C$ is a proportionality factor, and is most likely a function of the mean magnetic field since the fluctuations will change as the strength of the mean-field changes \citep[herein called \citetalias{Federrath2016c}]{Federrath2016c}. The key feature of this analysis is that from the induction equation we derived that there is a linear dependence between the magnetic field fluctuations ($\delta B$) and the turbulent velocity fluctuations ($\propto\M$). This is consistent with the strong-field model for the fluctuations proposed by \citetalias{Federrath2016c} (where $C = \Mao / 2)$, which will be discussed later in \S\ref{sec:1pointStats}. We note also that from Equation~(\ref{eq:induction4}), the ratio between $\delta B$ and $B_0$ is independent of $\M$, which we will also explore further. Next we discuss the simulations that we perform to test our models and explore the magnetic fluctuations.

\begin{figure*}
    \centering
    \includegraphics[width=\linewidth]{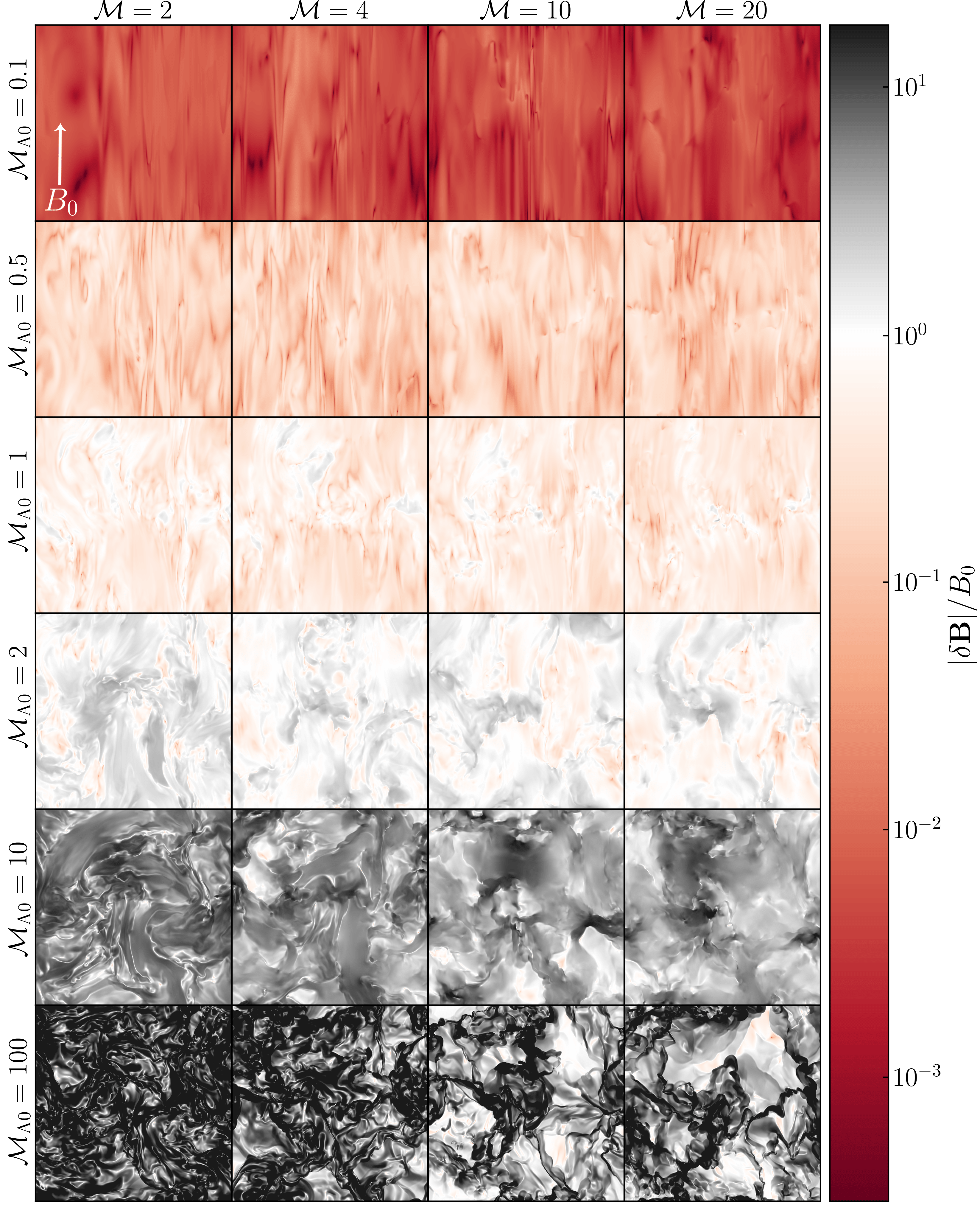}
    \caption{A slice of $|\delta\vecB{B}|$ compensated by $B_0$, at $t = 5\,T$, where $B_0$ is the magnitude of the mean of the magnetic field, which is orientated up the page, $\vecB{B}_0 = B_0 \hat{\vecB{z}}$. The turbulent Mach number increases from $\M = 2 - 20$, from left to right, and the Alfv\'enic Mach number of the mean magnetic field increases from $\Mao = 0.1 - 100$ from top to bottom, hence the simulation with the largest $B_0$ and $\M$ is in the top right corner, and the smallest $B_0$ and $\M$ in the bottom left corner. White indicates that $|\delta\vecB{B}| = B_0$, red $|\delta\vecB{B}| < B_0$ and black $|\delta\vecB{B}| > B_0$.}
    \label{fig:20panel}
\end{figure*}

\section{Numerical Simulations} \label{sec:Sims}

\begin{table*}
\caption{Main simulation parameters and derived quantities used throughout this study.}
\centering
\begin{tabular}{l r@{}l r@{}l c r@{}l r@{}l r@{}l}
\hline
\hline
& \multicolumn{2}{c}{Turbulence} & \multicolumn{3}{c}{Mean Magnetic Field} & \multicolumn{6}{c}{Fluctuating Magnetic Field} \\
\hline
\multicolumn{1}{c}{Simulation} & \multicolumn{2}{c}{$\M$} & \multicolumn{2}{c}{$\Mao$} & $B_0$ & \multicolumn{2}{c}{$\delta B_{\perp} \, [\mu\text{G}]$} & \multicolumn{2}{c}{$\delta B_{\parallel} \, [\mu\text{G}]$} & \multicolumn{2}{c}{$\delta B  \, [\mu\text{G}]$} \\
\multicolumn{1}{c}{ID} & \multicolumn{2}{c}{$(\pm1\sigma)$} & \multicolumn{2}{c}{$(\pm1\sigma)$} & $[\mu\text{G}]$ & \multicolumn{2}{c}{$(\pm 1\sigma)$} & \multicolumn{2}{c}{$(\pm 1\sigma)$} & \multicolumn{2}{c}{$(\pm 1\sigma)$} \\
\multicolumn{1}{c}{(1)} & \multicolumn{2}{c}{(2)} & \multicolumn{2}{c}{(3)} & (4) &\multicolumn{2}{c}{(5)} & \multicolumn{2}{c}{(6)} & \multicolumn{2}{c}{(7)} \\
\hline \\ 
\texttt{M2Ma100}       & 2.1\,    & $ \pm$ 0.1      & 103\,   & $ \pm$ 5        & 0.07      & 1.7\, & $\pm$ 0.1     & 1.2\, & $\pm$ 0.1     & 2.1\, & $\pm$ 0.2     \\[1em]
\texttt{M2Ma10}        & 1.80\,   & $ \pm$ 0.08     & 9.0\,   & $ \pm$ 0.8      & 0.71      & 3.8\, & $\pm$ 0.2     & 2.6\, & $\pm$ 0.1     & 4.6\, & $\pm$ 0.2     \\[1em]
\texttt{M2Ma2}         & 1.66\,   & $ \pm$ 0.05     & 1.7\,   & $ \pm$ 0.1      & 3.54      & 4.4\, & $\pm$ 0.4     & 2.7\, & $\pm$ 0.1     & 5.2\, & $\pm$ 0.4     \\[1em]
\texttt{M2Ma1}         & 2.0\,    & $ \pm$ 0.1      & 0.98\,  & $ \pm$ 0.07     & 7.09      & 3.5\, & $\pm$ 0.3     & 2.0\, & $\pm$ 0.1     & 4.1\, & $\pm$ 0.3     \\[1em]
\texttt{M2Ma0.5}       & 2.2\,    & $ \pm$ 0.2      & 0.54\,  & $ \pm$ 0.04     & 14.18      & 2.5\, & $\pm$ 0.3    & 1.5\, & $\pm$ 0.1     & 2.9\, & $\pm$ 0.3     \\[1em]
\texttt{M2Ma0.1}       & 2.6\,    & $ \pm$ 0.2      & 0.131\, & $ \pm$ 0.008    & 70.90      & 0.4\, & $\pm$ 0.04   & 0.8\, & $\pm$ 0.1     & 0.8\, & $\pm$ 0.1     \\[1em]
\texttt{M4Ma100}       & 4.0\,    & $ \pm$ 0.2      & 101\,   & $ \pm$ 4        & 0.14      & 2.6\, & $\pm$ 0.3     & 1.8\, & $\pm$ 0.3     & 3.1\, & $\pm$ 0.4     \\[1em]
\texttt{M4Ma10}        & 3.7\,    & $ \pm$ 0.1      & 9.2\,   & $ \pm$ 0.6      & 1.42      & 6.5\, & $\pm$ 0.4     & 4.4\, & $\pm$ 0.1     & 7.8\, & $\pm$ 0.4     \\[1em]
\texttt{M4Ma2}         & 3.5\,    & $ \pm$ 0.1      & 1.73\,  & $ \pm$ 0.07     & 7.09      & 8.3\, & $\pm$ 0.4     & 4.9\, & $\pm$ 0.2     & 9.6\, & $\pm$ 0.4     \\[1em]
\texttt{M4Ma1}         & 3.8\,    & $ \pm$ 0.3      & 0.95\,  & $ \pm$ 0.08     & 14.18      & 6.6\, & $\pm$ 0.6    & 3.6\, & $\pm$ 0.2     & 7.5\, & $\pm$ 0.7     \\[1em]
\texttt{M4Ma0.5}       & 4.4\,    & $ \pm$ 0.2      & 0.54\,  & $ \pm$ 0.03     & 28.36      & 4.7\, & $\pm$ 0.6    & 2.7\, & $\pm$ 0.3     & 5.4\, & $\pm$ 0.7    \\[1em]
\texttt{M4Ma0.1}       & 5.2\,    & $ \pm$ 0.4      & 0.13\,  & $ \pm$ 0.01     & 141.80     & 0.9\, & $\pm$ 0.2    & 1.0\, & $\pm$ 0.2     & 1.3\, & $\pm$ 0.3     \\[1em]
\texttt{M10Ma100}      & 10.0\,   & $ \pm$ 0.3      & 100\,   & $ \pm$ 3        & 0.35      & 5.9\, & $\pm$ 0.8     & 4.0\, & $\pm$ 0.6     & 7\, & $\pm$ 1     \\[1em]
\texttt{M10Ma10}       & 9.2\,    & $ \pm$ 0.4      & 9.2\,   & $ \pm$ 0.7      & 3.54      & 15.1\, & $\pm$ 0.6    & 9.9\, & $\pm$ 0.4     & 18\, & $\pm$ 1     \\[1em]
\texttt{M10Ma2}        & 9.0\,    & $ \pm$ 0.4      & 1.8\,   & $ \pm$ 0.1      & 17.72      & 19 \, & $\pm$ 1      & 11.7\, & $\pm$ 0.7    & 22\, & $\pm$ 1     \\[1em]
\texttt{M10Ma1}        & 9.3\,    & $ \pm$ 0.5      & 0.93\,  & $ \pm$  0.05    & 35.45      & 16 \, & $\pm$ 1      & 8.2\, & $\pm$ 0.6     & 18\, & $\pm$ 2     \\[1em]
\texttt{M10Ma0.5}      & 10.5\,   & $ \pm$ 0.4      & 0.52\,  & $ \pm$ 0.02     & 70.90      & 10 \, & $\pm$ 1      & 5.8\, & $\pm$ 0.8     & 12\, & $\pm$ 1    \\[1em]
\texttt{M10Ma0.1}      & 12\,     & $ \pm$ 1        & 0.125\, & $ \pm$ 0.006    & 354.49      & 2.3 \, & $\pm$ 0.3  & 2.5\, & $\pm$ 0.3     & 3.4\, & $\pm$ 0.5    \\[1em]
\texttt{M20Ma100}      & 20\,     & $ \pm$ 1        & 101\,   & $ \pm$ 4        & 0.71      & 9.9\, & $\pm$ 0.8     & 6.8\, & $\pm$ 0.6     & 12\, & $\pm$ 1     \\[1em]
\texttt{M20Ma10}       & 19\,     & $ \pm$ 1        & 9.3\,   & $ \pm$ 0.8      & 7.09      & 28 \, & $\pm$ 2       & 18.5\, & $\pm$ 0.8    & 34\, & $\pm$ 2     \\[1em]
\texttt{M20Ma2}        & 18\,     & $ \pm$ 1        & 1.8\,   & $ \pm$ 0.1      & 35.45      & 4.4\, & $\pm$ 0.3    & 2.7\, & $\pm$ 0.1     & 41\, & $\pm$ 2     \\[1em]
\texttt{M20Ma1}        & 19\,     & $ \pm$ 1        & 0.93\,  & $ \pm$ 0.03     & 70.90      & 31\, & $\pm$ 3       & 16\, & $\pm$ 1        & 35\, & $\pm$ 3     \\[1em]
\texttt{M20Ma0.5}      & 21\,     & $ \pm$ 1        & 0.53\,  & $ \pm$ 0.02     & 141.80      & 21\, & $\pm$ 3      & 11\, & $\pm$ 1        & 24\, & $\pm$ 3    \\[1em]
\texttt{M20Ma0.1}      & 24\,     & $ \pm$ 1        & 0.119\, & $ \pm$ 0.003    & 708.98      & 3.8\, & $\pm$ 0.4   & 4.1\, & $\pm$ 0.5     & 5.6\, & $\pm$ 0.6    \\[1em]
\hline 
\hline
\end{tabular} \\
\begin{tablenotes}
\item{\textit{\textbf{Notes:}} For each simulation we extract 51 $\vecB{B}$ realisations at 0.1$\, T$ intervals, where $T$ is the turbulent turnover time, between $5 \, T$ and $10 \, T$. All $1\sigma$ fluctuations listed are from the time-averaging over the $5\, T$. We show an example of the fluctuating component of the field in units of mean-field in Figure \ref{fig:20panel}. All simulations have $512^3$ grid cells. Column (1): the simulation ID. Column (2): the rms turbulent Mach number, $\M = \sigma_V / c_s$. Column (3): the Alfv\'en Mach number for the mean-$\vecB{B}$ component, $\vecB{B}_0$, $\Ma_0 = (2c_s\M\sqrt{\pi \rho_0}) / |\vecB{B}_0|$, where $\rho_0$ is the mean density, $c_s$ is the sound speed. Column (4): the magnitude of the mean magnetic field, $|\vecB{B}_0| = B_0$. Column (5): the fluctuating component of the magnetic field that is perpendicular to the mean magnetic field, where $B_{\perp} = B_x = B_y$, discussed in \S\ref{sec:1pointStats}, and plotted in Figure \ref{fig:perpAndpar}. Column (6): the same as column (5) but for the fluctuating component parallel to the mean magnetic field, $B_z = B_{\parallel}$. Column (7): the total magnitude of the fluctuations $\delta B = |\delta\vecB{B}| = \sqrt{\delta B_x^2 + \delta B_y^2 + \delta B_z^2}$.}
\end{tablenotes}
\label{tb:simtab}
\end{table*}

\subsection{Turbulent MHD Model}\label{sec:mhdModel}

    In this study we analyse the magnetic field fluctuations in 24 high-resolution, 3D turbulent, ideal magnetohydrodynamical (MHD) simulations with a non-zero mean magnetic field and an isothermal equation of state, $P = c_s^2 \rho$, where $P$ is the pressure, with $c_s$ normalised to 1. We use a modified version of the \textsc{flash} code, based on the public version 4.0.1 \citep{Fryxell2000,Dubey2008} to solve the MHD equations (\ref{eq:continuity}--\ref{eq:div0}) in a periodic box with dimensions $L^3$, on a uniform grid with a uniform resolution of $512^3$ grid cells, using the multi-wave, approximate Riemann solver framework described in \citet{Bouchut2010}, and implemented and tested in \citet{Waagan2011}. A comprehensive parameter set, including Mach number, mean magnetic fields, and derived quantities for each of the simulations is listed in Table~\ref{tb:simtab}.

\subsection{Turbulent Driving, Density and Velocity Fields}

    The turbulent acceleration field $\vecB{F}$ in Equation~(\ref{eq:momentum}) follows an Ornstein-Uhlenbeck (OU) process in time and is constructed such that we can control the mixture of solenoidal and purely compressive modes in $\vecB{F}$ (see \citealt{Federrath2008,Federrath2009,Federrath2010} for a detailed discussion of the turbulence driving). We choose to drive with a natural mixture of the two modes \citep{Federrath2010}. We isotropically drive in wavenumber space at $k \approx 2$, corresponding to real-space scales of $\ell_D \approx L/2$. The driving amplitude is centred on $k=2$ and falls off to zero with a parabolic spectrum towards $k=1$ on one side and $k=3$ on the other. Thus, only large scales are driven and the turbulence on smaller scales develops through the turbulent energy cascade driven from those large scales. The auto-correlation timescale of $\vecB{F}$ is equal to $T = L /(2c_s \M)$, where $L$ is the system scale, hence we use the auto-correlation timescale of $\vecB{F}$ to set the turbulent turnover timescale for the desired turbulent Mach number on $L/2$ for each simulation. We vary the sonic Mach number between $\M =$ 2 and 20, encompassing the range of observed $\M$ values for the turbulent interstellar medium (e.g., \citealt{Schneider2013,Federrath2016b,Orkisz2017,Beattie2019b}). The initial velocity field is set to $\vecB{v}(x,y,z,t=0)=0$ and the density field $\rho(x,y,z,t=0)=1$, i.e., the mean density, $\rho_0=1$. We run the simulations for $10\,T$, where $T$ is the eddy turnover time. We note that the physical evolution of these systems is fully described by the dimensionless $\M$ and $\Mao$ numbers, and all dimensional quantities, such as $L$, $\rho$, $v$, $B$, etc., can be scaled arbitrarily, as long as their chosen values leave $\M$ and $\Mao$ unchanged for a given simulation model.
    
    \begin{figure*}
        \centering
        \includegraphics[width=0.73\linewidth]{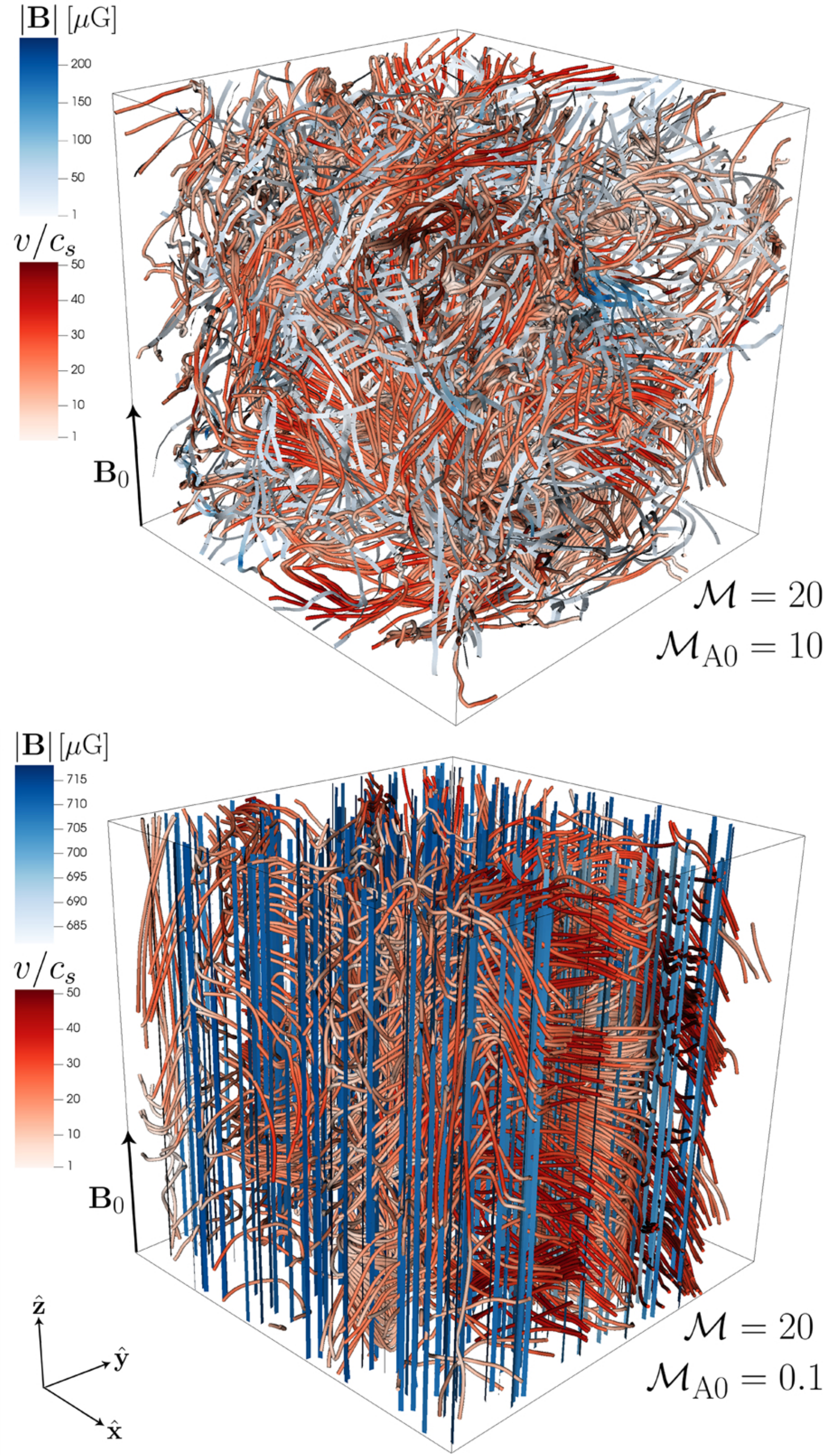}
        \caption{The 3D geometry of magnetic field (blue) and velocity streamlines (red) for the $\M=20$, $\Mao = 10$ (top) and $\M=20$, $\Mao = 0.1$ (bottom) simulations at $t = 5\,T$. We see isotropic magnetic and velocity structure for the super-Alfv\'enic turbulence, and highly-anisotropic magnetic fields and vortex creation in the velocity streamlines for the sub-Alfv\'enic turbulence. In incompressible flows the sub-Alfv\'enic turbulence becomes quasi two-dimensional, but in our compressible supersonic simulations, shocks form perpendicular to the vortices \citep{Beattie2020}, which sustain parallel fluctuations in the magnetic field.}
        \label{fig:3Dplots}
    \end{figure*}

\subsection{Magnetic Fields} \label{sec:Bfields}

    The initial magnetic field, $\vecB{B}(x,y,z)$ at $t=0$, in Equations~(\mbox{\ref{eq:momentum}--\ref{eq:div0}}) is a uniform field with field lines threaded through the $\hat{\vecB{z}}$ direction of the simulations. The total magnetic field is, as given in Equations~\ref{eq:fluctuate} and~\ref{eq:magField} and reiterated here for clarity,
    \begin{align}
    \vecB{B}(t) = B_0 \vecB{\hat{z}} + \delta\vecB{B}(t).
    \end{align}
    Since we work in units of sound speed for velocity, and in units of $\rho_0$ for density, we chose to scale the units of the magnetic field for a typical molecular cloud density (hydrogen number density of $n_{\text{H}} = 10^3\,\mathrm{cm}^{-3}$) and temperature ($10\,\mathrm{K}$) corresponding to a sound speed of $c_\mathrm{s}=0.2\,\mathrm{km\,s}^{-1}$, to
    \begin{align}
    \vecB{B} = \vecB{B}_{\text{sim}}\left(\frac{c_s}{0.2 \, \text{km\,s}^{-1}}\right) \left( \frac{n_{\text{H}}}{10^3\,\text{cm}^{-3} } \right)^{1/2},
    \end{align}
    where $\vecB{B}_{\text{sim}}$ is the magnetic field in simulation units, such that all magnetic fields are in units of $\mu\text{G}$, as previously done in \citet{Mocz2018}, for example. We list all of the magnetic field components in Table \ref{tb:simtab}. Note that in our simulations $0.01 \lesssim B_0/(\mu\text{G}) \lesssim 1000$, providing us with $\sim$ five orders of magnitude to explore for the value of $B_0$ (column~4 in Table~\ref{tb:simtab}). This parameter space encompasses the wide variety of molecular clouds that one might observe.
    
    The fluctuating component of the field evolves self-consistently from the MHD equations. However, we set $B_0$, and by using the definition of the Alfv\'en velocity (defined previously in reference to Equation~\ref{eq:Ma}) and the turbulent Mach number, we can set the target $\Mao$,
    \begin{equation}\label{eq:meanField}
    \Mao = 2c_s\sqrt{\pi\rho_0}\M/B_0. 
    \end{equation}
    We vary this value for each of the 24 simulations, spanning $\Mao = 0.1 - 100$, encompassing very weak and strong mean magnetic fields. We extract 51 time realisations of the $B_x$, $B_y$ and $B_z$ components of the total magnetic field between $5 \leq t/T \leq 10$ to ensure that the turbulence is in a statistically stationary state \citep{Federrath2009,Price2010}.
    
    We show a slice of the magnitude of $\delta\vecB{B}$ in units of $B_0$, for each of the 24 simulations at $t = 5\,T$, in Figure~\ref{fig:20panel}. Red colours in the plot show $|\delta\vecB{B}| < B_0$, white $|\delta\vecB{B}| = B_0$ and black $|\delta\vecB{B}| > B_0$. In the sub-Alfv\'enic regime we see $|\delta\vecB{B}| < B_0$, in the super-Alfv\'enic regime $|\delta\vecB{B}| > B_0$ and $\Mao \approx 1$ marks the transition between the two. Using the magnetic field components we construct the variables $ B_{\parallel} = B_z$ and $ B_{\perp} = B_x = B_y$\footnote{Note that we are defining these magnetic field quantities within the global mean magnetic field frame, and not the local mean magnetic field frame. See \citet{Cho2002} for the difference between the two frames.}. We create PDFs for the two new magnetic field variables, and average them over $5 - 10\,T$. In \S\ref{sec:1pointStats} we discuss these PDFs in detail, but first we will discuss one of the main predictions from our fluctuation model in  Equation~(\ref{eq:induction2}), namely that it is primarily the velocity gradients parallel to the mean-field that create compressive modes in the velocity field.

    \begin{figure*}
        \centering
        \includegraphics[width=\linewidth]{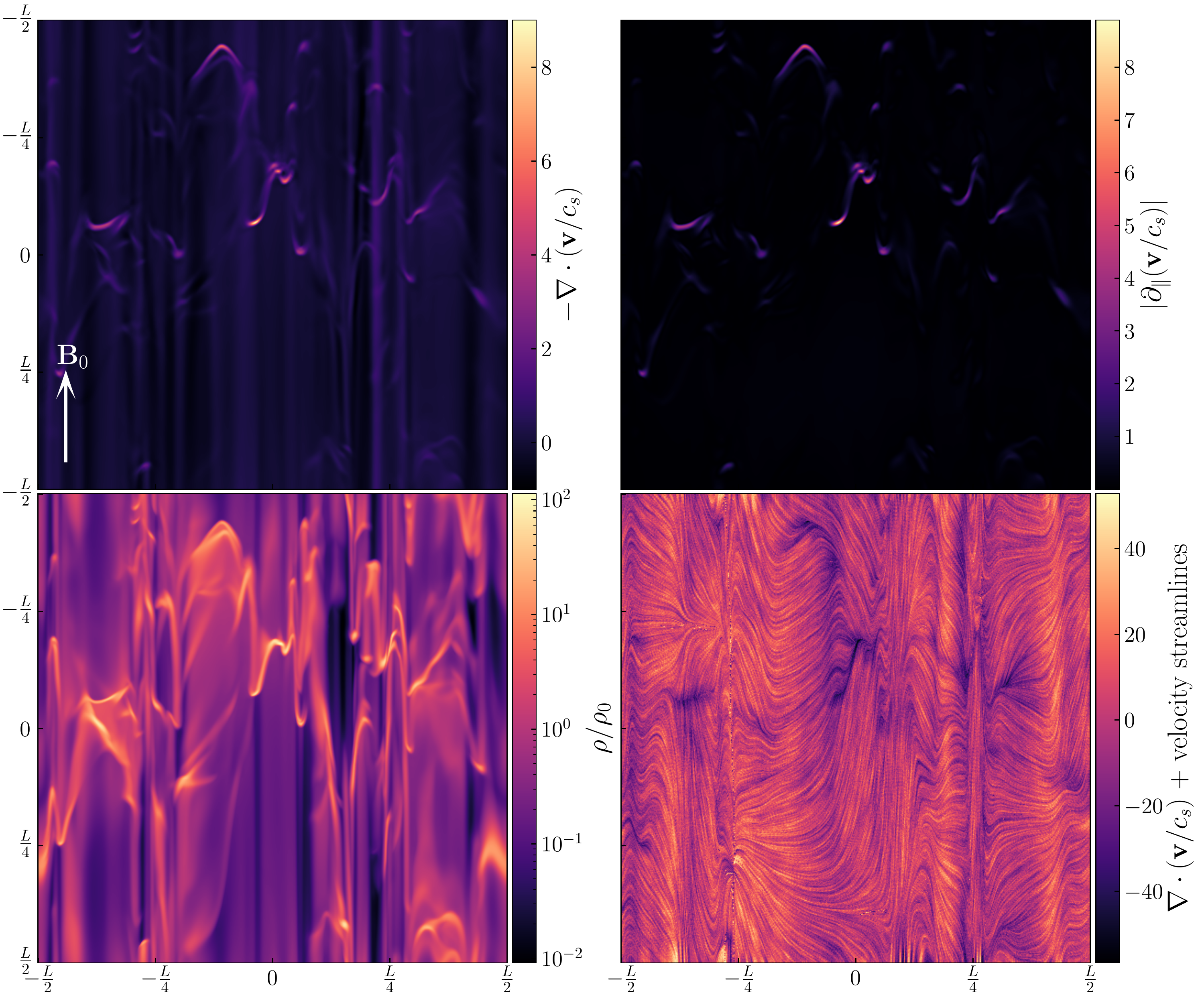}
        \caption{\textbf{Top-left:} The convergence of the velocity for a single $xz$ slice from the \texttt{M20MA0.1} simulation at $t = 5\,T$. The direction of the mean magnetic field is shown in the bottom left corner. We see highly $-\nabla \cdot \vecB{v} > 0$ (converging) structures forming perpendicular to $\vecB{B}_0$. Fast magnetosonic waves can also be seen travelling as compressive structures perpendicular to $\vecB{B}_0$ \citep{Tritsis2016}. \textbf{Top-right:} The magnitude of the velocity gradient along the magnetic field, $|\partial_{\parallel}\vecB{v}|$. Equation~(\ref{eq:induction2.1}) predicts that this plot and the magnitude of the convergence should be the same as $\partial_t\delta B \rightarrow 0$. This is indeed true, and the only significant difference between the two plots is the fast magnetosonic waves, which will disappear as $B_0 \gg \delta B$. \textbf{Bottom-left:} The same as the top-left plot, but for a density slice, shown in units of mean density, $\rho_0$. We see that the $-\nabla \cdot \vecB{v} > 0$ structures in the top panel correspond to hydrodynamical shocks \citep{Mocz2018,Beattie2020}, forming from \textit{mostly} parallel velocity streams in the flow. \textbf{Bottom-right:} Velocity streamlines integrated through the divergence of the velocity, showing how parallel velocity channels feed into high-density filaments, creating the velocity gradients parallel to the magnetic field.}
        \label{fig:divV}
    \end{figure*}
    
    \begin{figure*}
        \centering
        \includegraphics[width=\linewidth]{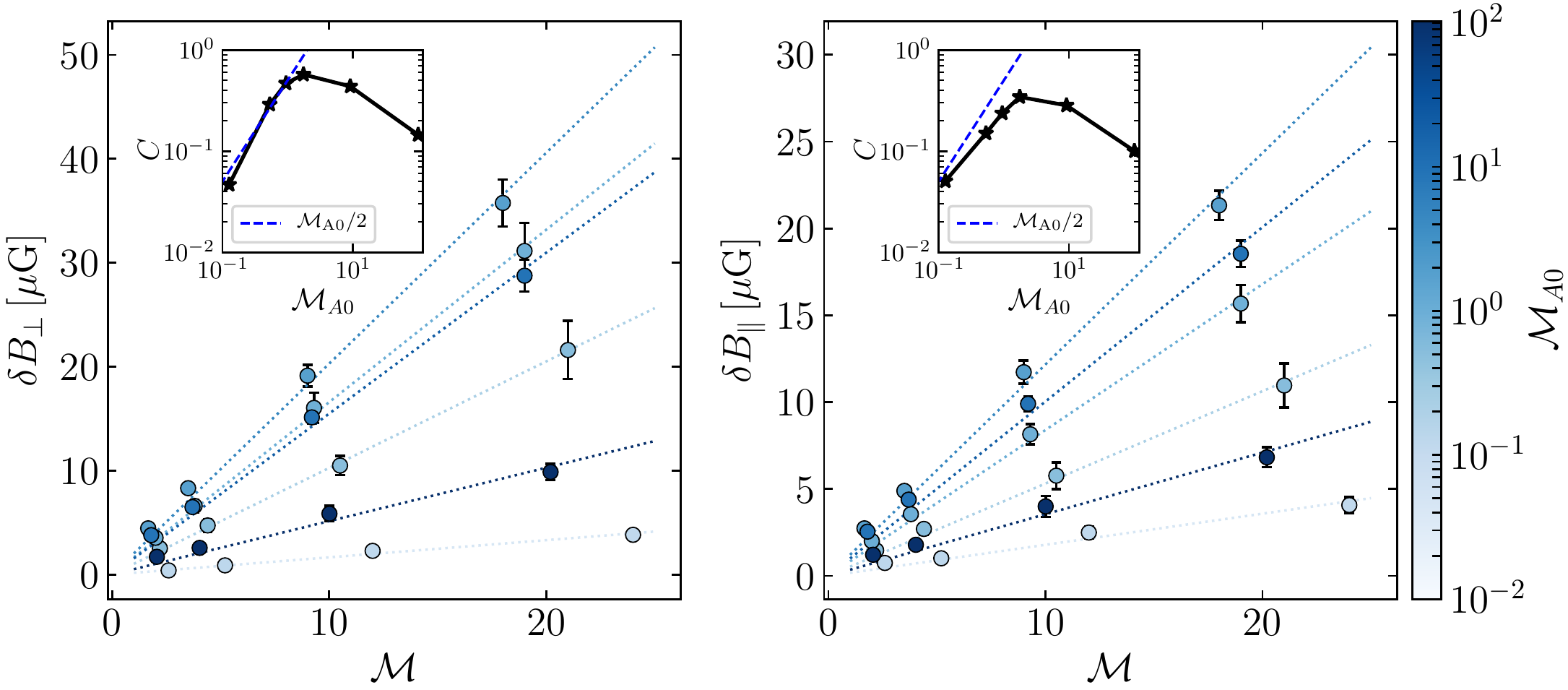}
        \caption{\textbf{Left:} The perpendicular magnetic field fluctuations (Equation~\ref{eq:BperpFluc}) as a function of $\M$, coloured by $\Mao$. The model fits, $\delta B_{\perp} = 2 c_s \sqrt{\pi\rho_0} C \M$, (Equation~\ref{eq:beattieModel}) are shown with dashed lines, also coloured by $\Mao$. The $C$ parameter from the model is shown in the inset plot, for each different $\Mao$. We show a blue, dashed line for $C = \Mao/2$, which is the strong magnetic field model derived in \citetalias{Federrath2016c} and written in terms of $\M$ and $\Mao$ in  Equation~(\ref{eq:fedMod}). \textbf{Right:} The same as the left panel but for the parallel magnetic field fluctuations. Both plots reveal that there is a linear dependence of the fluctuations on $\M$, as predicted by our compressible quasi-static model (Equation~\ref{eq:induction2}), and a close-to-linear dependence for $\Mao < 1$, i.e., for sub-Alfv\'enic mean-field flows.}
        \label{fig:perpAndpar}
    \end{figure*}
    
    \begin{figure}
        \centering
        \includegraphics[width=\linewidth]{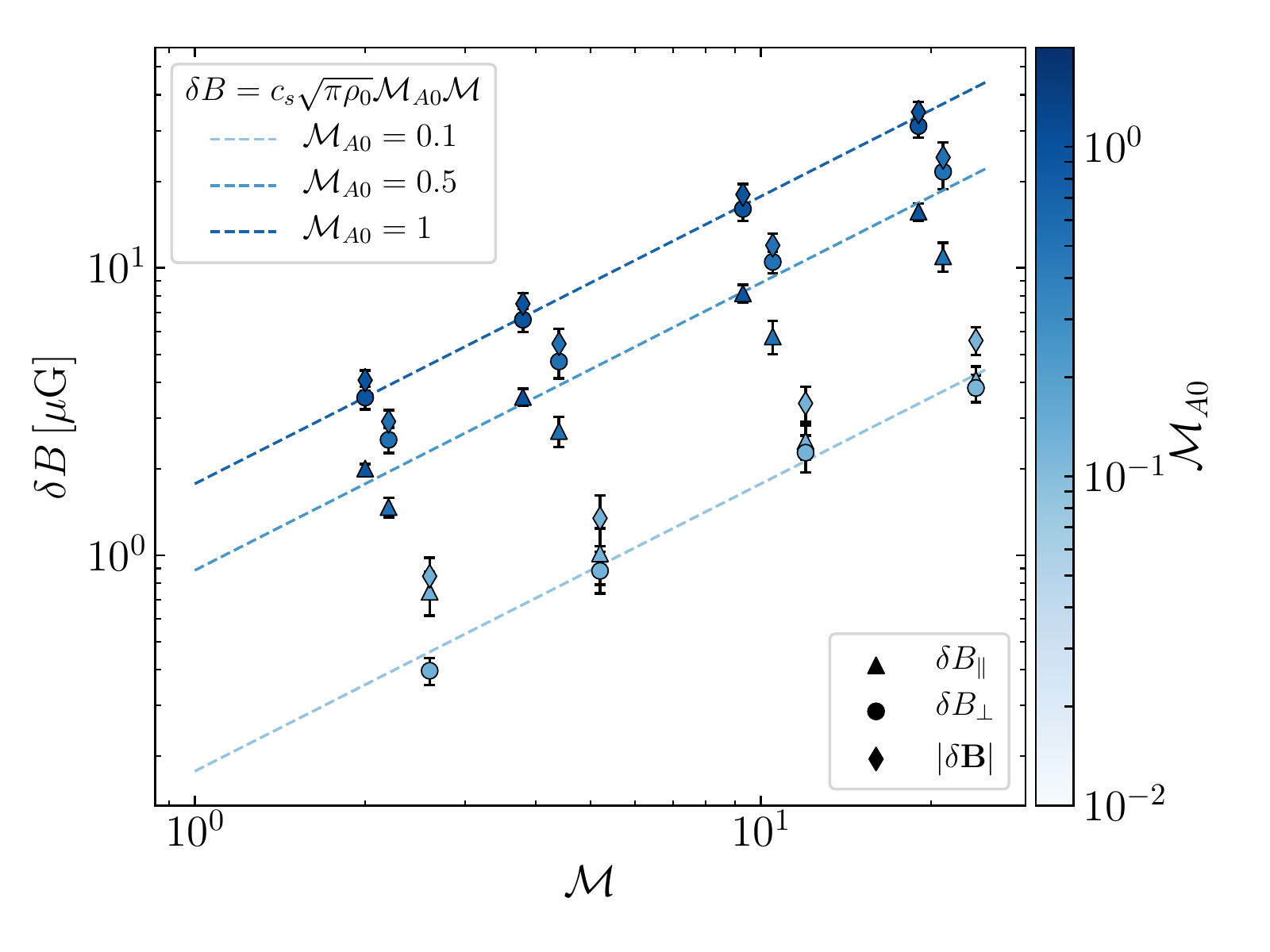}
        \caption{The parallel (shown with triangles), perpendicular (shown with circles) and total (shown with diamonds) fluctuations of the magnetic field as a function of $\M$, coloured by $\Mao$. We plot the modified \citetalias{Federrath2016c} magnetic field fluctuation model,  Equation~(\ref{eq:fedMod}), using dashed lines for $\Mao = 0.1$, $0.5$ and $1$. We show only the sub-Alfv\'enic simulations where the modified \citetalias{Federrath2016c} model is valid. }
        \label{fig:federrathModel}
    \end{figure}

\section{Compressible flows along the mean magnetic field}\label{sec:CompressFlow}

    In \S\ref{sec:theory} we showed how one could construct a compressive quasi-static model for the magnetic field fluctuations (Equation~\ref{eq:induction2}) and that in the strong mean-field regime the model reduces to a simple statement about compressibility and velocity gradients along $B_0$, shown in  Equation~(\ref{eq:induction2.1}). This is a significant result for astrophysical flows, since both strong, coherent mean magnetic fields and velocity gradients perpendicular to filamentary structures are measured in observations \citep{Palmeirim2013,Beuther2015,Federrath2016,Simajiri2019,Chen2020}. Here we show some direct, qualitative evidence to support Equation~(\ref{eq:induction2.1}). 
    
    We show a single time realisation of the full 3D geometry of the magnetic and velocity (scaled by $c_s$) fields in Figure \ref{fig:3Dplots} for the \texttt{M20MA10} (top) and  \texttt{M20MA0.1} (bottom) simulations, where we refer to Table~\ref{tb:simtab} for the respective labels. \texttt{M20MA10} shows an example of the tangled, isotropic magnetic fields that we have in the high $\Mao$ simulations. The magnetic field is dominated by $\delta\vecB{B}$, because the mean-field component is too weak to impart any systematic, ordered structure on either the velocity or total magnetic field. On the other hand, the \texttt{M20MA0.1} simulation shows a highly-ordered, anisotropic magnetic field, dominated by $\vecB{B}_0$. Many of the velocity streamlines form $\sim L/2$ eddies perpendicular to the magnetic field lines. If all of the streamlines were organised into eddies the flow would be statistically the same along $\vecB{B}_0$, hence becoming quasi two-dimensional. However, we see some parallel velocity streams developing towards the left corner of the plot. To explore this further we take an $xz$-slice through this same time realisation.
    
    We plot the convergence of the $xz$ slice of the velocity field for the $\texttt{M20MA0.1}$ simulation in the top-left panel of Figure~\ref{fig:divV}. Hydrodynamical shocks that form perpendicular to the mean-field (see \citealt{Mocz2018} for the shock-jump conditions, showing that they are hydrodynamical) can be clearly identified as $-\nabla \cdot \vecB{v} > 0$ structures, i.e., converging flows, by comparing the convergence with the density structures in the bottom-left panel of Figure~\ref{fig:divV}. Tracing the velocity streamlines, shown in the bottom-right panel, we can see the most significant shocks are coupled with a strong velocity channels along the mean magnetic field. Furthermore, probing the magnitude of the velocity gradient along $B_0$, shown in the top-left panel, shows a strong correlation between the $-\nabla \cdot \vecB{v} > 0$ structures and the velocity gradient, consistent with Equation~(\ref{eq:induction2.1}). The only difference between the panels is the $\nabla \cdot \vecB{v}$ signatures of fast magnetosonic waves that form perpendicular to the field in the convergence plot. However, these waves will cease as the mean-field becomes strong and magnetic tension locks them in place and then $|\nabla \cdot \vecB{v}|$ will indeed equal $|\partial_{\parallel} \vecB{v}|$. To summarise, what we show here is that as $B_0 \gg \delta B$ parallel velocity gradients cause compressive modes in the velocity field, which cause strong hydrodynamic shocks to form across the field. 
    
    Indeed, this MHD phenomenon may contribute to the anisotropic cloud collapse mechanism that is seen in sub-Alfv\'enic supersonic turbulence, simulations and (most likely) observations \citep{Tritsis2015,Mocz2017} since velocity streams could be responsible for accreting material into dense, star-forming filaments. Compressive modes that steepen velocity gradients could even contribute to coherent velocity gradients and flows along and across the filamentary structures in the clouds, like observed in star-forming and infrared dark clouds \citep{Palmeirim2013,Beuther2015,Simajiri2019} and investigated in great detail in \citet{Chen2020,Chen2020a}. \citet{Chen2020a} suggest that the velocity gradients are caused by gravitational accretion onto a self-gravitating filament. Here we add that compressive modes in highly-magnetised gases are coupled with velocity gradients along the direction of the mean magnetic field, which are perpendicular to the densest filaments, and parallel to striations \citep{Beattie2020}, and could also contribute to the accretion process. Furthermore, \citet{Molina2012} found, when measuring the density dispersion for sub-Alfv\'enic turbulence, that more compressive modes appeared relative to the compressions induced by turbulent forcing alone (i.e. the $b$ parameter that is an input in the simulations). This may also be explained by the emergence of more compressive modes when the turbulence is in this highly-magnetised state. Certainly this result warrants detailed investigation in future studies and we will be further exploring these velocity structures in an ensemble of higher-resolution simulations in a forthcoming study on the two-point velocity statistics of anisotropic, supersonic MHD turbulence.
    
    To conclude this section we comment on the velocity gradient method, which is summarised in \citet{Hu2019}, and used to measure the $\Ma$ values that we report for the Gould Belt in \S\ref{sec:intro}. This method has been significantly developed in the literature \citep[and references therein]{Casanova2017,Yuen2017,Yuen2017b,Lazarian2018b} and assumes that velocity gradients are perpendicular to magnetic fields, which may be true for incompressible $\Mao < 1$ flows (i.e., through the two-dimensionalisation due to the exponentially decaying parallel magnetic field fluctuations; \citealt{Verma2017}), but is not be strictly true for compressible flows, as indicated in  Equation~(\ref{eq:induction2.1}) and the discussion above. \citet{Yuen2017b} attribute co-alignment of velocity and magnetic field vectors to indicate gravitational collapse. We show that this need not be the case, and for supersonic flows in a sub-Alfv\'enic mean-field regime, one can indeed find strong velocity gradients parallel to the mean magnetic field, without the presence of self-gravity. Next we discuss the magnetic field PDFs, and how the magnetic field fluctuations depend upon $\M$ and $\Mao$.

\section{Structure of the Magnetic Field} \label{sec:1pointStats}

In this section we provide a detailed study of the magnetic field structure for each of our 24 simulations, over a wide range of $\M$ and $\Mao$. We first explore the magnitude and anisotropy of the fluctuations, followed by mean-normalised fluctuations and finally the morphology and intermittency of the magnetic field PDFs.

\subsection{Magnetic field fluctuations}

    We construct the PDFs for both the $ B_{\parallel}$ and $ B_{\perp}$ components of the magnetic field and average them over the time range $5 - 10\,T$. The first moment and second moment of the PDFs have a physical interpretation. For example, for the $ B_{\parallel}$ PDF the first moment is,
    \begin{align} \label{eq:B0dis}
        \Exp{  B_{\parallel}} = B_0 = \int \d{  B_{\parallel}} \,   B_{\parallel} p(  B_{\parallel}) ,
    \end{align}
    where $p(  B_{\parallel})$ is the PDF for the parallel component of the magnetic field, describes the mean-field value from  Equation~(\ref{eq:meanField}). Likewise, the variance of the field is 
    \begin{align}
       \Exp{  B_{\parallel}^2} - \Exp{  B_{\parallel}}^2 = \delta   B_{\parallel}^2 = \int \d{  B_{\parallel}} \, \left(  B_{\parallel} - B_0 \right)^2 p(  B_{\parallel}),
    \end{align}
    which means the standard deviation of the field is exactly $\delta    B_{\parallel}$. Since the mean-field only has a $z$ component, $\Exp{ B_{\perp}} = 0$, and hence the second moment of the PDF is exactly equal to the fluctuations squared,
    \begin{align}\label{eq:BperpFluc}
       \Exp{ B_{\perp}^2} = \delta  B_{\perp}^2 = \int \d{ B_{\perp}} \,  B_{\perp}  p( B_{\perp}).
    \end{align}
    Less important in our study is the first moment of these PDFs, since we set this as an input parameter in our simulations, as described in \S\ref{sec:Sims}. Hence we focus our analysis on the standard deviation of these distributions, which tells us information about the magnitude of the total fluctuations across all (perpendicular or parallel to the mean guide field) length scales in the simulations. For a scale-dependent analysis of the fluctuations one would need to calculate a two-point statistic, which we plan to explore in detail, in a future study.

\subsubsection{$\delta  B_{\perp}$ and $\delta   B_{\parallel}$ fluctuations}\label{sec:flucamplitudes}

    In Figure \ref{fig:perpAndpar} we show the perpendicular and parallel fluctuation amplitudes of the magnetic field as a function of $\M$, coloured by $\Mao$. Both types of fluctuations show a linear dependence in $\M$, as predicted by our order-of-magnitude estimate of the fluctuations in  Equation~(\ref{eq:beattieModel}). Conceptually, the $\M$ dependence can be explained by the turbulent motions contributing to tangling, twisting and perturbing the magnetic fields. More quantitatively, in our quasi-static model for the fluctuations, shown in Equation~(\ref{eq:induction2}), each of the RHS terms depend linearly on the velocity field, regardless of the relation between $B_0$ and $\delta B$, hence the linear dependence upon $\M$ is a somewhat universal feature for the amplitude of the fluctuations. For a constant $\Mao$, we fit Equation~(\ref{eq:beattieModel}),
    \begin{equation}
        \delta B = 2c_s\sqrt{\rho_0 \pi} C \M,
    \end{equation}
    where $C$ is the slope parameter. Note that we need not include an offset in the fits, since when $\M \rightarrow 0$ there is no source for the magnetic field fluctuations, hence $\delta B \rightarrow 0$. Clearly $C$ is a function of $\Mao$, which is shown by how the slopes change for different $\Mao$ in Figure \ref{fig:perpAndpar}, and plotted in the inset figure for each panel. For $\Mao \lesssim 1$ we see an approximately linear dependence, $C \approx \Mao/2$, for both types of fluctuations, until reaching a turnover at $\Mao \approx 1-2$, where reducing the mean-field strength decreases the $C$ parameter. This is because as $|\vecB{B}_0|$ shrinks the total magnetic field energy strength also shrinks, and thus the magnitude of the fluctuations decreases, at least until the turbulent small-scale dynamo mechanism can exponentially grow the fluctuations \citep[and references therein]{Batchelor1950,Kazantsev1968,Schekochihin2002,Federrath2014, Seta2015, Seta2020}, which is a regime we do not consider in this study (but is studied in \citetalias{Federrath2016c}). Next we explain the linear dependence of $\Mao$ in the strong-field regime.

\subsubsection{Strong-field Model}

    Using the definition of the Alfv\'en Mach number, $\Mao = V / V_{A0}$, the turbulent Mach number, $\M = V / c_s$, and the Alfv\'en velocity, $V_{A0} = B_0 / \sqrt{4\pi\rho_0}$ one can show that the strong-field \citetalias{Federrath2016c} model (Equation~\ref{eq:fed2016mod}) can be rewritten in terms of $\M$,
    \begin{align}\label{eq:fedMod}
        \delta B = c_s\sqrt{\pi\rho_0} \Mao \M.
    \end{align}
    This modified version of the \citetalias{Federrath2016c} model shows a linear dependence in $\M$, in agreement with our model derived from the induction equation, and linear in $\Mao$. Hence, the $C$ obtained is analytical in this regime, $C = \Mao/2$. This means, as suggested in the derivation of Equation~(\ref{eq:beattieModel}), that $C$ can be thought of as the Alfv\'enic control parameter, i.e., it changes based on the type of Alfv\'enic turbulence, which is shown clearly in the inset plots of Figure \ref{fig:perpAndpar}. 
    
    We plot Equation~(\ref{eq:fedMod}) on the magnitudes of the perpendicular, parallel and total fluctuations of the magnetic field in Figure \ref{fig:federrathModel}, where we show only data from the trans- to sub-Alfv\'enic regimes, since this is the regime where the strong-field model is valid. In general, we find that the model captures the nature of the fluctuations well, tracking either values between $ B_{\perp}$ and $  B_{\parallel}$ in the sub-Alfv\'enic simulations, or very close to the total fluctuations in the trans-Alfv\'enic simulation, $\Mao \sim 1$. Plotting both perpendicular and parallel fluctuations in Figure \ref{fig:federrathModel} reveals the difference between their magnitudes due to the anisotropy inherent to MHD turbulence with a mean-field present \citep{Goldreich1995,Lazarian1999,Cho2000,Cho2003,Boldyrev2006,Kowal2007,Burkhart2014,Burkhart2015}. In the following, we explore the anisotropy of the fluctuations.
    
    \begin{figure}
        \centering
        \includegraphics[width=\linewidth]{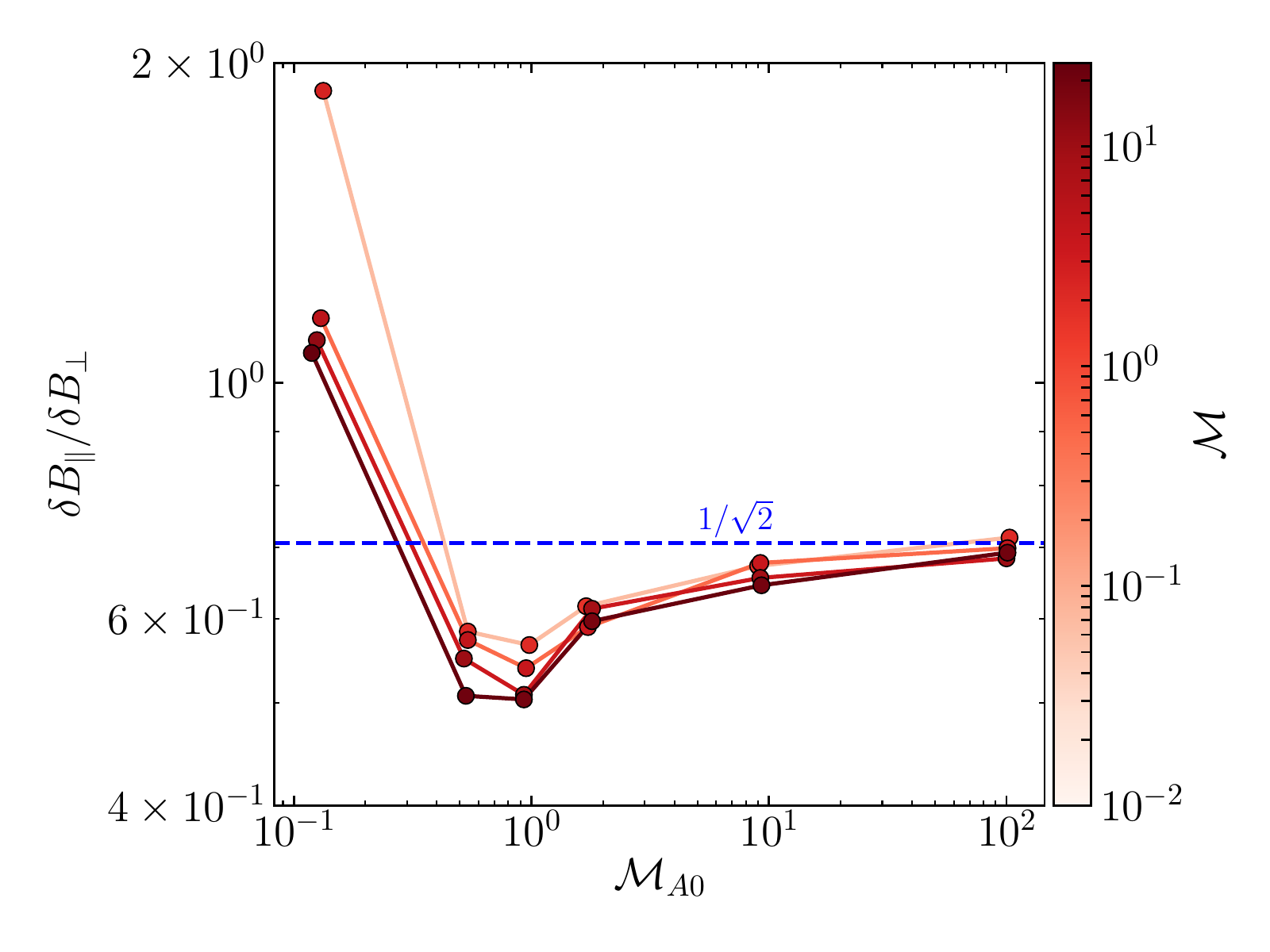}
        \caption{The ratio between $\delta  B_{\parallel}$ and $\delta B_{\perp}$ as a function of $\Mao$, coloured by $\M$. The anisotropy in the fluctuations is largest for small $\Mao$, and disappears for large $\Mao$, where $\delta B_{\parallel}/\delta B_{\perp} = 1/\sqrt{2}$ is completely isotropic, indicated by the blue, dashed line.}
        \label{fig:anisotropy}
    \end{figure}

\subsection{Anisotropy of the magnetic fluctuations}

    In Figure \ref{fig:anisotropy} we show the ratio between the $ B_{\parallel}$ and $ B_{\perp}$ fluctuations. Since $\delta B_{\perp} = \sqrt{\delta B_x^2 + \delta B_y^2}$ when $\delta B_x = \delta B_y = \delta B_{\parallel}$ the ratio $\delta B_{\parallel} / \delta B_{\perp} = 1/\sqrt{2}$. For $\Mao \gtrsim 2$ we see the ratio monotonically approaching the $1/\sqrt{2}$ isotropic limit. We can understand this from our model in Equation~(\ref{eq:induction2}). When $\delta B \gg B_0$ we can disregard the $\mathcal{O}(B_0)$ terms in Equation~(\ref{eq:induction2}), 
    \begin{align} \label{eq:induction_isotropic}
        \frac{D}{Dt}\delta\vecB{B} = & (\delta\vecB{B} \cdot \nabla)\vecB{v}  - \delta\vecB{B}(\nabla \cdot \vecB{v}),
    \end{align}
    where $\frac{D}{Dt} \equiv \left(\frac{\partial }{\partial t} + \vecB{v}\cdot\nabla\right)$, which no longer has the anisotropic term, $B_0 \partial_{\parallel} \vecB{v}$. This means that each of the components of $\delta \vecB{B}$ are affected by the advection and compression equally, causing isotropic fluctuations, without any preferential direction (which we visualised in the top panel of Figure \ref{fig:3Dplots}). However, in the opposite case, where $B_0 \gg \delta B$, the $B_0 \partial_{\parallel} \vecB{v}$ becomes large and we measure $\delta B_{\parallel}$ up to a factor 2 larger than $\delta B_{\perp}$. This is because the $B_0 \partial_{\parallel} \vecB{v}$ from our quasi-static model causes strong compressions across the field, which increases the magnetic field fluctuations along the mean-field direction, as shown in Figure \ref{fig:divV}, and discussed in detail in \S\ref{sec:theory} and \S\ref{sec:CompressFlow}. The impact that these shocks have on the parallel fluctuations is quite phenomenal, and we see a significant growth in the absolute value of the anisotropy between $\Mao = 0.5$, which has stronger $ B_{\perp}$ fluctuations, and $\Mao = 0.1$.
    
    An important feature seen from Figure \ref{fig:anisotropy} is that the ratio between magnetic field fluctuations divides out most of the $\M$ dependence, which we can see from each of the different $\M$ simulations falling approximately upon each other. This suggests that the anisotropy in the magnetic field fluctuations are controlled mostly by $\Mao$, although some $\M$ dependence is certainly present in the $\Mao = 0.1$ simulations. This motivates another avenue of enquiry, the $B_0$ normalised fluctuations, which should also be $\M$ independent according to Equation~(\ref{eq:induction4}). We explore this in the next section.
    
    \begin{figure}
        \centering
    \includegraphics[width=\linewidth]{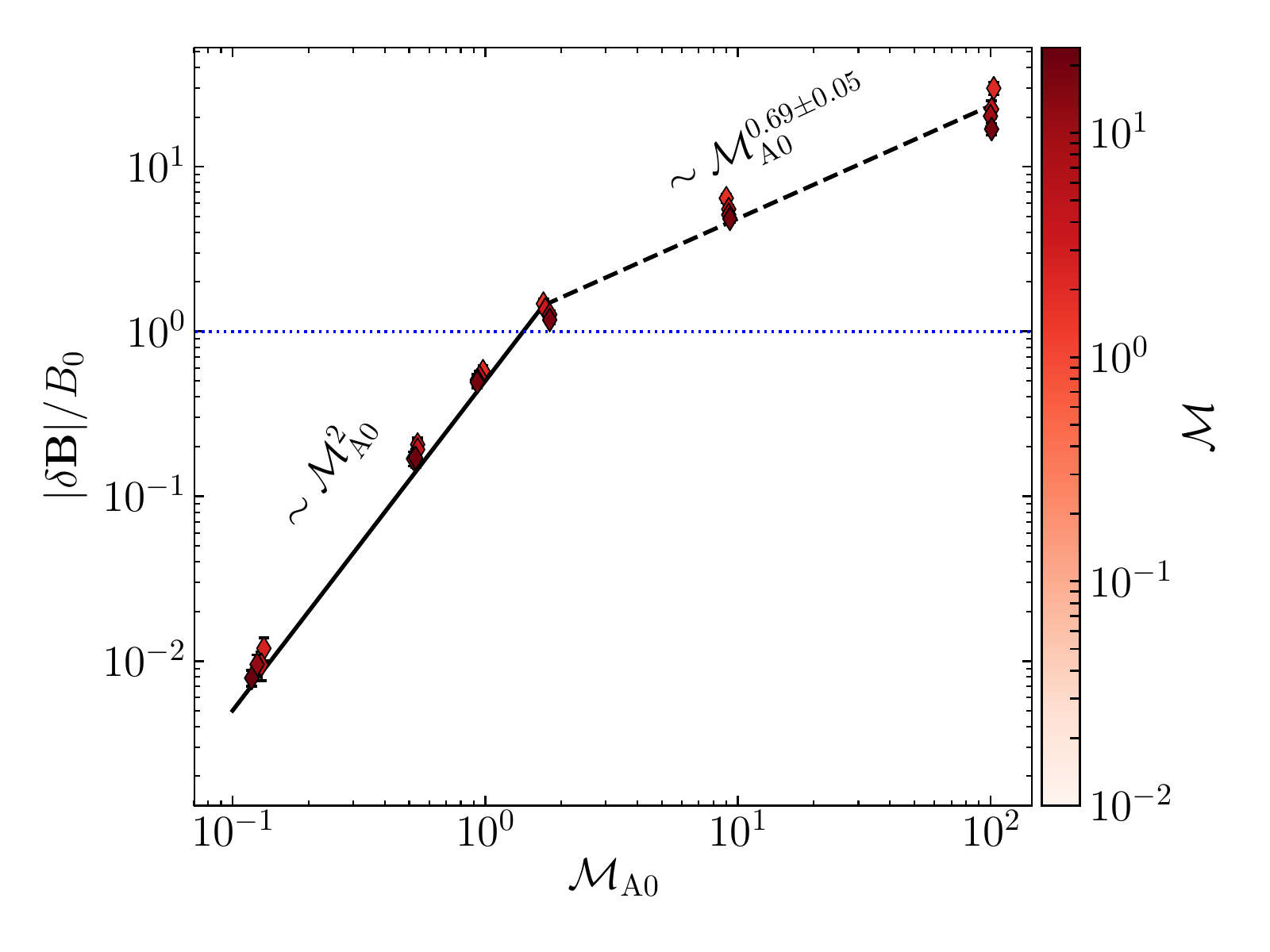}
        \caption{The mean-field normalised fluctuations, $\delta B / B_0$, as a function of $\Mao$, coloured by $\M$. The vertical blue line indicates when $(\delta B / B_0) = 1$. The solid black line is the strong-field model, Equation~( \ref{eq:fedMod}), and the dashed black line is our fit, $(\delta B / B_0) = \MaO{0.69 \pm 0.05}$, fit on $2 \leq \Mao \leq 100$ data.}
        \label{fig:weightedFluc}
    \end{figure} 

\subsection{Mean-field normalised fluctuations}\label{sec:meanFieldWeighted}

    We show the mean-field normalised fluctuations in Figure \ref{fig:weightedFluc} as a function of $\Mao$, coloured by $\M$, for the total magnetic field fluctuations. We do not find a large difference in the two fluctuating components of the field (see Appendix \ref{appendix:meanFieldWeighted}), so we focus our analysis on the total fluctuations. As we saw in the Figure \ref{fig:anisotropy}, there is only a very weak $\M$ dependence left, which we show explicitly in Figure~\ref{fig:machDepend_meanfield} in the Appendix, which is the same as Figure \ref{fig:perpAndpar}, but for the mean normalised quantities). This plot shows variations in $\M$, regardless of the value for $\Mao$, will not influence $\delta B / B_0$. We can see why this is the case in our compressible quasi-static model, since $\delta B / B_0 = C(\Mao) \Mao$ as shown in Equations~\mbox{\ref{eq:induction4} -- \ref{eq:beattieModel}}.
    
    One of the key features of Figure \ref{fig:weightedFluc} is the distinct kink at $\delta B \approx B_0$ (indicated by the horizontal blue line in Figure \ref{fig:weightedFluc}), which happens at $\Mao \approx 2$. The same transition can be seen in Figure \ref{fig:20panel}, where fluctuations are $\mathcal{O}(B_0)$, and are becoming more isotropic in nature, which is shown in Figure \ref{fig:anisotropy}. Clearly this marks a critical transition for the magnetic field evolution, where all of the terms in our quasi-static model play a role, so very little can be deduced using our model in this transition region. Instead, we propose a semi-analytical model that encapsulates both the $\delta B / B_0 < 1$ and $\delta B / B_0 > 1$ regimes,
    \begin{align}
        \frac{\delta B}{B_0} = & C \Mao, \\
        C = & 
    \left\{\begin{matrix}
    \Mao/2, \, \text{for}\;\delta B/B_0 < 1,
    \\[0.5em] \alpha\MaO{\beta}, \, \text{for}\;\delta B / B_0 \geq 1,
    \end{matrix}\right.
    \end{align}
    where $\alpha$ and $\beta$ are fit parameters. For $\delta B/B_0 < 1$ we use the \citetalias{Federrath2016c} model, but since no analytical model is known in the $\delta B / B_0 \geq 1$ regime, and our compressible quasi-static model does not give much insight, we instead directly fit to the data. Using a linear least-squares fit we determine $\alpha \approx 1$ and $1 + \beta = 0.69 \pm 0.05$. This is somewhat consistent with the qualitative model for the ``intermediate regime" in \citetalias{Federrath2016c} ($\delta B \sim B_0^{1/3} \implies (\delta B /B_0) \sim B_0^{-2/3} \sim \MaO{2/3}$), but here we explicitly show that the mean-field weighted fluctuations are independent of $\M$, regardless of the ratio between $\delta B$ and $B_0$. Hence the relation is,
    \begin{align} \label{eq:deltaB/B0_relation}
        \frac{\delta B}{B_0}  = & 
    \left\{\begin{matrix}
    \MaO{2}/2, \, \text{for}\; \delta B/B_0 < 1,\\[0.5em] 
    \MaO{0.69 \pm 0.05}, \, \text{for}\; \delta B / B_0 \geq 1,
    \end{matrix}\right.
    \end{align}
    which we show with the solid and dashed black line in Figure \ref{fig:weightedFluc}, respectively. Indeed, we expect that our model will hold until the dynamo regime is reached as $\delta B / B_0 \gg 1$ and for all $\M$, making it a universal scaling relation for anisotropic, supersonic MHD turbulence. This is the key result from this investigation. 
    
    The relative scatter around our relation is clearly set by $\M$, and becomes larger as the mean field weakens. Calculating the standard deviation for each of the fixed $\Mao$ values in Figure \ref{fig:weightedFluc} we find that the relative scatter changes systematically with $\Mao$, ranging between $\approx 5 - 20 \, \%$. Now that we have discussed in great detail the absolute values, the anisotropy and the mean-field weighted $\delta B$ we move on to give a deeper, and more complete analysis of the magnetic field PDFs. 
    \begin{figure*}
        \centering
        \includegraphics[width=\linewidth]{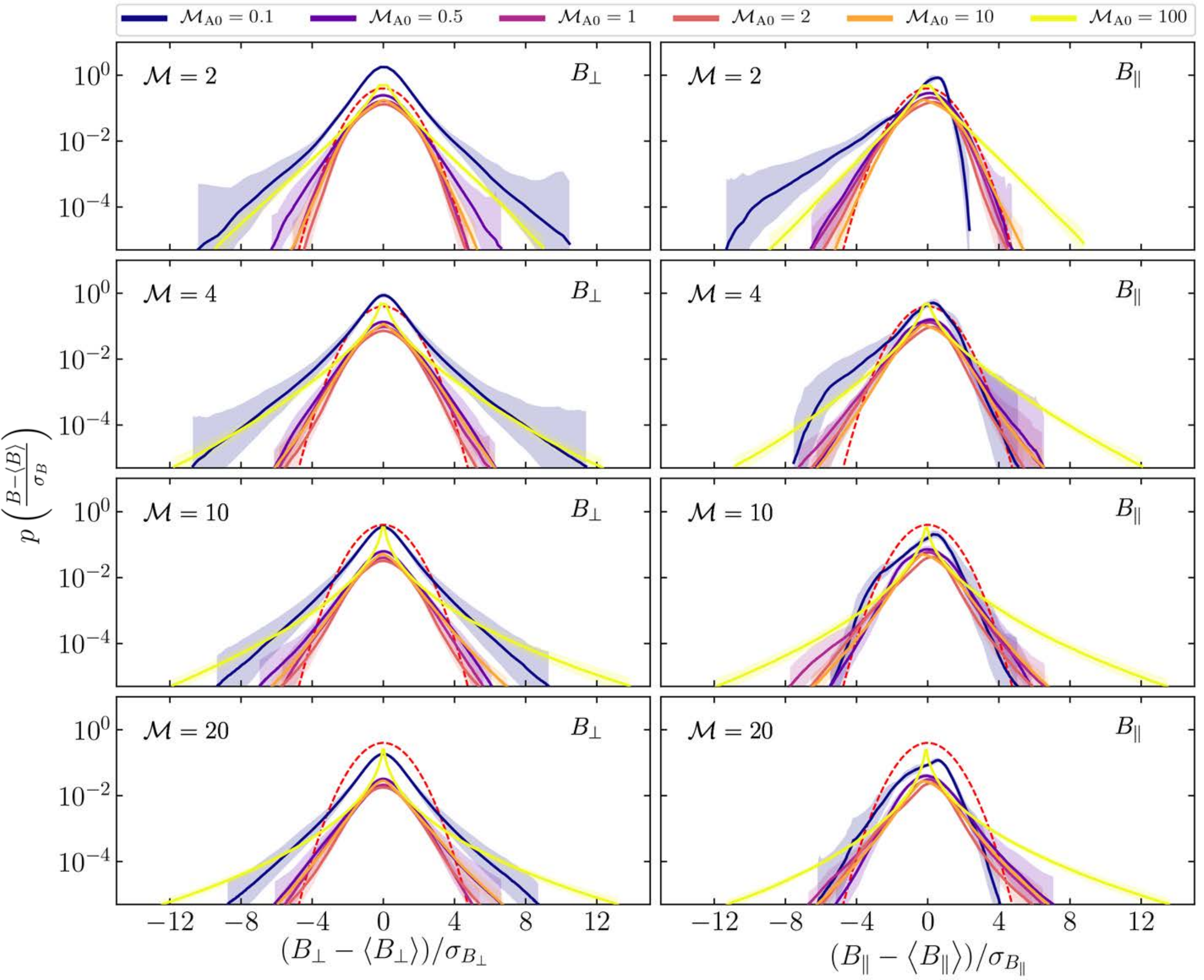}
        \caption{Variance ($=1$) and mean ($=0$) normalised PDFs of $ B_{\perp}$ (left column) and $ B_{\parallel}$ (right column), for each of the 24 simulations revealing the morphology of the distributions. Each panel shows simulations with different $\M$ values, and within each panel shows the ensemble with different $\Mao$ values, with the legend shown at the top. We normalise by $\sigma_{B} = \delta B$ values for each of the distributions because it makes it easier to compare the shapes of the distributions without focusing on the amplitudes of the fluctuations, which are discussed in detail in \S\ref{sec:flucamplitudes}. We plot a standard Gaussian, $\mathcal{N}(0,1)$, in red for comparative purposes. $1\sigma$ fluctuations are shown with the bands around the $5 - 10\,T$ time average.}
        \label{fig:BPerpBparDis}
    \end{figure*}

\subsection{Magnetic field distribution and intermittency}\label{sec:morphology}

    Magnetic field fluctuations are highly intermittent \citep[specifically see \S3 in \citealt{Seta2020}]{Roberto2007,Seta2020}, i.e., the fluctuations cannot be aptly described by the variance of the magnetic field distribution that we have described in detail in the previous sections. In this section, however, we systematically describe the shape and features of the distributions for the $ B_{\perp}$ and $ B_{\parallel}$ components of the turbulent magnetic field, for each of the simulations. We show the time-averaged distributions in Figure \ref{fig:BPerpBparDis}. Note that we plot the distributions of the normalised variable, $(B - \Exp{B}) / \sigma_B$, which enforces that all of the distributions have a mean of zero and variance of one. This allows us to compare the shape of the distributions without being obscured by the absolute magnitude of the magnetic field fluctuations, which were studied in \S\ref{sec:flucamplitudes}. We now consider the $ B_{\perp}$ and $ B_{\parallel}$ PDFs in the following two subsections. 

\subsubsection{$ B_{\perp}$ PDF}

    In the left column of Figure \ref{fig:BPerpBparDis} we show the perpendicular magnetic field distributions. The distributions are reasonably symmetric, for all $\Mao$ and $\M$. Indeed, as described earlier in \S\ref{sec:Bfields}, this is because the Lorentz force acts symmetrically about the mean magnetic field, defining an axis of symmetry for the fluctuations \citep{Cho2002}. We plot a standardised Normal distribution, $\mathcal{N}(0,1)$, shown in dashed-red in each of the panels to compare the distributions with purely Gaussian fluctuations. The most Gaussian fluctuations are found in the $\M = 2$ simulations where $\delta B \gtrsim B_0$, i.e., for $\Mao \approx 2 - 10$, as shown in Figure \ref{fig:weightedFluc}. This marks the transition between the low-$\Mao$ and $\delta B \gg B_0$ flows. The key difference between $\mathcal{N}(0,1)$ and the other distributions are the long, extended tails and peaked mode, shown in both the low- and high-$\Mao$ simulations for $\M = 2$, and then present in all of the distributions for $\M \geq 4$. This is a signature of intermittency. 
    
    In the low-$\Mao$ simulations, where the magnetic field is extremely strong and under large amounts of tension, which scales as $1/\MaO{2}$, the intermittent tails of the distributions can be associated with only large magnetic field perturbations. These could be from strong hydrodynamical shocks, which are intermittent events in the velocity field. Shock number densities (in a given volume, for example) are a function of $\M$, with more shocks and with more power in each shock, being created for higher values of $\M$ \citep{Gotoh1994,Girimaji1995,Beattie2020b}. This is why as the $\M$ increases, we see more intermittency in all of the distributions, regardless of $\Mao$. In the high-$\Mao$ simulations, like the $\Mao = 100$ ensemble, the mean-field is extremely weak compared to the fluctuations. Hence these distributions are only composed of the fluctuating magnetic field, which is purely intermittent in nature. Next we discuss the $ B_{\parallel}$ PDF.

\begin{figure}
    \centering
    \includegraphics[width=\linewidth]{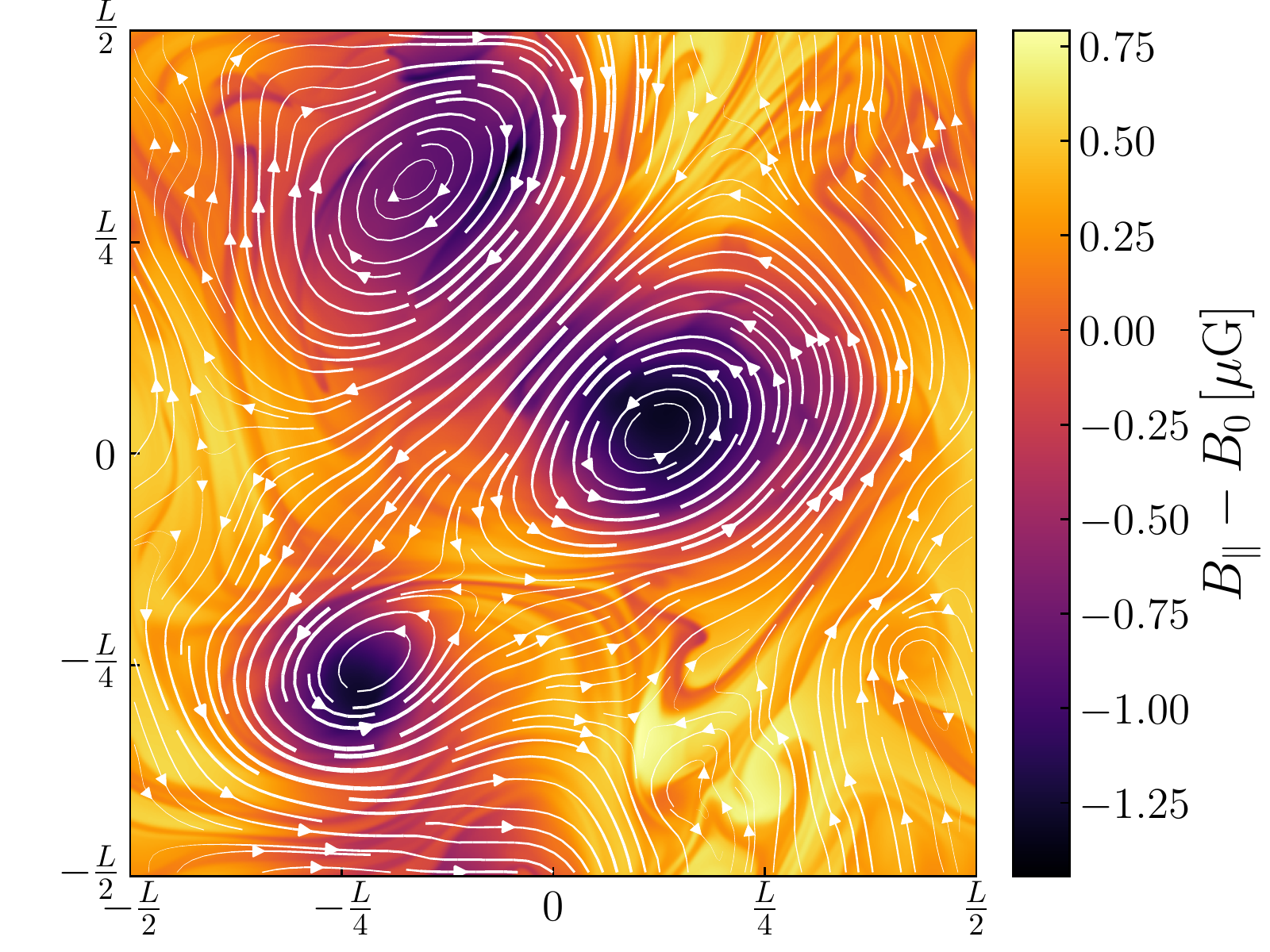}
    \caption{Magnetic eddies shown in an $xy$ slice of the $B_{\parallel}$ component of the magnetic field for the \texttt{M2MA0.1} simulation. Velocity streamlines, with the thickness of the line proportional to the magnitude of the velocity, are overlaid on the plot to reveal the eddy structure. We subtract the mean field to reveal the regions of negative $\delta B_{\parallel}$, corresponding to the extended log-linear tails in the $B_{\parallel}$ PDFs from Figure \ref{fig:BPerpBparDis}, discussed in \S\ref{sec:BparDis}.}
    \label{fig:magneticEddies}
\end{figure}

\subsubsection{$ B_{\parallel}$ PDF}\label{sec:BparDis}

    In the right column of Figure \ref{fig:BPerpBparDis} we show the parallel magnetic field PDFs. In the high-$\Mao$ regime, the fluctuations are mostly symmetrical about the mean, similar to the $ B_{\perp}$ PDFs described previously. However, for the sub-Alfv\'enic flows we see extended $\log$-linear tails developing in the negative values of the PDFs, most obviously shown in the top, right panel, for the $\texttt{M2MA0.1}$ simulation. This extended tail corresponds to values of the total magnetic field, which are less than $B_0$, i.e., $B_{\parallel} < B_0$, where $B_{\parallel} = B_0 + \delta B_{\parallel}$, are the parallel components of the magnetic field from Equation~(\ref{eq:magField}). Clearly this implies that preferentially $\delta B_{\parallel}$ opposes the mean field, reducing the total magnetic field where the fluctuation is present, i.e., in these local regions, the parallel components of the vector that describes the fluctuation must have the form $B_{\parallel} = B_0 - \delta B_{\parallel}$. The tail is most prominent in the morphology of the low-$\M$ simulations.
   
    By creating a mask that reveals just the magnetic field values in the tail of the $B_{\parallel}$ PDF we find that eddies are responsible for creating localised magnetic voids in the field. We show an example of the magnetised eddies in the \texttt{M2MA0.1} simulation, where we plot $B_{\parallel} - B_0$ for an $xy$ slice through the magnetic field, overlaid with velocity streamlines in Figure~\ref{fig:magneticEddies}. The eddies form low-density and magnetic pressure regions in the fluid, and with a fixed $B_0$ this means that the fluctuating component of the field must oppose the mean-field to reduce the local magnetic pressure, giving rise to the extended $B_{\parallel} < B_0$ tails in the PDFs. However, as the magnetic fields become more isotropic with increasing $\M$ (Figure \ref{fig:anisotropy}) the tails become less prominent as the low pressure regions are mixed through the flow.
    
    The remaining structure in the distribution is then most likely due to the turbulent fluctuations, which we showed in \S\ref{sec:flucamplitudes} are linear in $\M$. Hence for large $\M$ values we mostly see the velocity interactions, and the interaction between $\delta  B_{\perp}$ and $\delta  B_{\parallel}$ no longer dominates the morphology of the PDFs.

\section{Summary and Key Findings}\label{sec:conclusion}

    In this study, we explore the amplitude of magnetic field fluctuations that are perpendicular, $\delta B_{\perp}$, and parallel, $\delta  B_{\parallel}$, to the mean magnetic field, $\vecB{B}_0$, in an ensemble of 24 simulations, across a large range of Alfv\'enic mean-field Mach numbers, $\Mao = 0.1 - 100$ and root-mean-squared turbulent Mach numbers, $\M = 2 - 20$, encompassing realistic values for molecular clouds, the birthplaces of stars. First, we derive a new compressible quasi-static fluctuation model, which we use to explain much of the phenomena discussed in this study and show how, even in the strong mean magnetic field regime the flow remains three-dimensional. Next we explore how the nature of the fluctuations perpendicular and parallel to the mean magnetic field change as a function of $\M$ and $\Mao$, and generalise an analytical model for the fluctuations in the strong mean-field regime. This is followed by an investigation of the anisotropy between $\delta B_{\perp}$ and  $\delta  B_{\parallel}$, and the mean-field normalised fluctuations. Finally, we explore the shape of the  probability density functions of magnetic fluctuations. We summarise the key findings below:
    \begin{itemize}
        \item We derive a compressible quasi-static model for the magnetic field fluctuations, shown in Equation~(\ref{eq:induction2}) with full derivation in Appendix \ref{appendix:induction}, which predicts that for sub-Alfv\'enic flows, compressions (e.g., shocks) in the velocity field are associated with gradients along the mean magnetic field, $$|\nabla\cdot\vecB{v}| = |\partial_{\parallel} \vecB{v}|,$$ which could be a contributing factor for anisotropic cloud collapse \citep[etc.]{Tritsis2015,Mocz2017} and coherent velocity structures perpendicular to the principle axis of filaments, observed in real molecular clouds \citep{Chen2020}. We use the non-dimensional form of the equation to predict that the magnitude of the magnetic field fluctuations are linear in $\M$, $$ \delta B = 2c_s \sqrt{\pi\rho_0} C(\Mao) \M,$$ where $c_s$ is the sound speed, $\rho_0$ is the mean density and $C$ is the proportionality factor, which changes between the $B_0 \gg \delta B$ and $B_0 \ll \delta B$ regimes. We show in Figure \ref{fig:perpAndpar} that the linear dependence predicted for all $\Mao$ holds true. \\[1em]
        \item We plot the velocity divergence and velocity streamlines for one of the $\Mao = 0.1$ simulations in Figure \ref{fig:divV}, revealing the compressive modes coupled with parallel velocity gradients, $|\nabla\cdot\vecB{v}| = |\partial_{\parallel} \vecB{v}|$. We show that, even in the absence of self-gravity, shocks that form perpendicular to the mean magnetic field turn into converging flows that are accreted upon by coherent parallel velocity channels, consistent with our compressible quasi-static model. \\[1em]
        \item Our new fluctuation model reduces to the \citet{Federrath2016c} analytical model for strong-field magnetic field fluctuations by setting $C = \Mao / 2$. We further rewrite the \citet{Federrath2016c} model to incorporate the explicit dependence upon $\M$, $$\delta B = c_s\sqrt{\pi\rho_0}\Mao\M,$$ and we show in Figure \ref{fig:federrathModel} that it is in good agreement with the measured fluctuations for the sub-Alfv\'enic simulations, across all $\M$. \\[1em]
        \item We find the ratio between the $\delta  B_{\perp}$ and $\delta  B_{\parallel}$ components of the magnetic field are similar for different $\M$, but do depend upon $\Mao$ (shown in Figure \ref{fig:anisotropy}). As $\Mao$ gets larger the fluctuations become isotropic, but for $\Mao \sim 0.1$ $\delta  B_{\parallel} \sim 2 \delta  B_{\perp}$. This is because the strong velocity gradients along the field, when $B_0 \gg \delta B$, leads to large values of $\delta  B_{\parallel}$. This is a key distinguishing feature from incompressible MHD turbulence with a strong mean magnetic field, where parallel fluctuations exponentially decay and the flow becomes quasi two-dimensional \citep{Alexakis2011,Verma2017}. We show that the flow remains very much three-dimensional for compressible MHD turbulence. \\[1em]
        \item We show that the mean-field normalised fluctuations, $\delta B / B_0$, are independent of $\M$ in Figure \ref{fig:weightedFluc}. We propose a semi-analytic model using the \citet{Federrath2016c} model in the strong-field regime and a least-squares fit in the weak-field regime,
        $$
        \frac{\delta B}{B_0}  = 
        \left\{\begin{matrix}
        \Mao^2/2, \, \delta B/B_0 < 1,\\[0.5em] 
        \MaO{0.69 \pm 0.05}, \, \delta B / B_0 \geq 1,
        \end{matrix}\right.
        $$
        which is valid until the turbulence reaches the dynamo growth regime, $(\delta B/B_0) \gg 1$, and independent of $\M$, making it a universal feature of anisotropic, supersonic MHD turbulence. \\[1em] 
        \item We calculate the time-averaged ($5 - 10\,T$, where $T$ is the large-scale turbulent eddy turnover time) $ B_{\perp}$ and $ B_{\parallel}$ distributions, shown in Figures \ref{fig:BPerpBparDis}, and discuss their morphology in \S\ref{sec:morphology}. We find distinct signatures of intermittency in most of the distributions and an extended tail into the negative values for the $ B_{\parallel}$ distribution. This corresponds to $ B_{\parallel}$ fluctuations opposing the mean-field, which we show is caused by eddies that cause local, low-magnetised pressure regions in the fluid.
    \end{itemize}
    
    Data analysis and visualisation software used in this study: \textsc{numpy} \citep{Oliphant2006}, \textsc{matplotlib} \citep{Hunter2007}, \textsc{cython} \citep{Behnel2011}, \textsc{visit} \citep{Childs2012}, \textsc{scipy} \citep{Virtanen2020}.
    
\section*{Acknowledgements}
    We thank the reviewer, Pierre Lesaffre, for providing a thorough and constructive review of our study. We thank the Australian National University and Research School of Astronomy and Astrophysics for the support in place for staff and students during the COVID19 isolation, which is when most of this study was written. J.~R.~B. acknowledges funding from the Australian National University, specifically the Deakin PhD and Dean's Higher Degree Research (theoretical physics) Scholarships. C.~F.~acknowledges funding provided by the Australian Research Council (Discovery Project DP170100603, and Future Fellowship FT180100495), and the Australia-Germany Joint Research Cooperation Scheme (UA-DAAD). We further acknowledge high-performance computing resources provided by the Leibniz Rechenzentrum and the Gauss Centre for Supercomputing (grants~pr32lo, pr48pi and GCS Large-scale project~10391), the Partnership for Advanced Computing in Europe (PRACE grant pr89mu), the Australian National Computational Infrastructure (grant~ek9), and the Pawsey Supercomputing Centre with funding from the Australian Government and the Government of Western Australia, in the framework of the National Computational Merit Allocation Scheme and the ANU Allocation Scheme. The simulation software FLASH was in part developed by the DOE-supported Flash Centre for Computational Science at the University of Chicago. 

\section*{Data Availability}

The data underlying this article will be shared on reasonable request to the corresponding author.


\bibliographystyle{mnras.bst}
\bibliography{March2020.bib} 



\appendix
\section{The compressible quasi-static fluctuations model}\label{appendix:induction}
    The general MHD induction equation is
    \begin{equation}
        \frac{\partial \vecB{B}}{\partial t} = \nabla \times (\vecB{v} \times \vecB{B}),
    \end{equation}
    but with a magnetic field decomposition that we show in Equation~(\ref{eq:magField}) it is possible to simplify the equation, and gain some physical intuition for the fluctuating component of $\vecB{B}$. First we write $\vecB{B}(t) = \vecB{B}_0 + \delta\vecB{B}(t)$,
    \begin{equation}
        \frac{\partial \delta\vecB{B}}{\partial t} = \nabla \times (\vecB{v} \times [ \vecB{B}_0 + \delta\vecB{B}]),
    \end{equation}
    since $\partial_t \vecB{B_0} = 0$. Expanding the cross product twice we find,
    \begin{align} \label{eq:expandInduc}
        \frac{\partial \delta\vecB{B}}{\partial t} = \nabla \times (\vecB{v} \times \vecB{B}_0) + \nabla \times (\vecB{v} \times \delta\vecB{B}). 
    \end{align}
    Next we use the identity $\nabla \times ( \vecB{A} \times \vecB{B}) = \vecB{A} (\nabla \cdot \vecB{B}) - \vecB{B}(\nabla \cdot \vecB{A}) + (\vecB{B}\cdot \nabla)\vecB{A} - (\vecB{A}\cdot \nabla)\vecB{B}$ to expand each of the two terms. The first term is,
    \begin{align}
        \nabla \times (\vecB{v} \times \vecB{B}_0) = & \vecB{v} (\nabla \cdot \vecB{B}_0) - \vecB{B}_0(\nabla \cdot \vecB{v}) \nonumber\\
        & + (\vecB{B}_0\cdot \nabla)\vecB{v} - (\vecB{v}\cdot \nabla)\vecB{B}_0,
    \end{align}
    which we can simplify because the magnetic field is divergence free, and because $\partial_{x_i}\vecB{B}_0 = 0$, hence,
    \begin{align}
        \nabla \times (\vecB{v} \times \vecB{B}_0) = - \vecB{B}_0(\nabla \cdot \vecB{v}) + (\vecB{B}_0\cdot \nabla)\vecB{v}.
    \end{align}
    Since $\vecB{B}_0 = B_0 \hat{\vecB{z}}$ the second term can be written in terms of just the velocity gradient in the $z$ direction,
    \begin{align} \label{eq:expandB0}
        \nabla \times (\vecB{v} \times \vecB{B}_0) = B_0 \partial_z \vecB{v} - \vecB{B}_0(\nabla \cdot \vecB{v}).
    \end{align}
    Now we turn our attention to the fluctuating term in Equation~(\ref{eq:expandInduc}). Using the same identity as we used for the mean-field term and immediately dropping the $\nabla \cdot \delta\vecB{B}$ term we find,
    \begin{align} \label{eq:expandBfluc}
        \nabla \times (\vecB{v} \times \delta\vecB{B}) = -\delta\vecB{B}(\nabla \cdot \vecB{v}) + (\delta\vecB{B}\cdot \nabla)\vecB{v} - (\vecB{v}\cdot \nabla)\delta\vecB{B}.
    \end{align}
    Now combing the two Equation~(\ref{eq:expandB0}) and (\ref{eq:expandBfluc}) to construct Equation~(\ref{eq:expandInduc}),
    \begin{align}
        \nabla \times (\vecB{v} \times \vecB{B}_0) + & \nabla \times (\vecB{v} \times \delta\vecB{B}) = \nonumber \\
        & B_0 \partial_z \vecB{v} - \vecB{B}_0(\nabla \cdot \vecB{v}) + \nonumber \\
        & (\delta\vecB{B}\cdot \nabla)\vecB{v} - (\vecB{v}\cdot \nabla)\delta\vecB{B} -\delta\vecB{B}(\nabla \cdot \vecB{v}),
    \end{align}
    which can be simplified to reveal,
    \begin{align}
        \frac{\partial \delta\vecB{B}}{\partial t} + (\vecB{v}\cdot \nabla)\delta\vecB{B} =
        & B_0 \partial_z \vecB{v} - \vecB{B}(\nabla \cdot \vecB{v}) + (\delta\vecB{B}\cdot \nabla)\vecB{v},
    \end{align}
    or more succinctly, using $\frac{D}{Dt} = \partial_t + \vecB{v}\cdot\nabla$ as the Lagrangian derivative, i.e. the derivative in the frame co-moving with the fluid,
    \begin{align}
        \frac{D\delta\vecB{B}}{D t} =
        & B_0 \partial_z \vecB{v} - \vecB{B}(\nabla \cdot \vecB{v}) + (\delta\vecB{B}\cdot \nabla)\vecB{v},
    \end{align}
    The first term, $B_0 \partial_z \vecB{v}$, (we use the notation $B_0 \partial_\parallel \vecB{v}$ in the main text) tells us that the fluctuations change when the velocity gradient along the mean magnetic field changes. This term encodes the anisotropy of the fluctuations into the induction equation. Thus, when $B_0$ increases (or equivalently $\Mao$ decreases) the anisotropy in fluctuations increases and the fluctuations become more isotropic. This is also demonstrated via results from our simulations in Figure~\ref{fig:anisotropy}. The second term, $\vecB{B}(\nabla \cdot \vecB{v})$, tells us that the change in the fluctuations scale with the compression of the velocity field weighted by the magnetic field. The last term, $(\delta\vecB{B}\cdot \nabla)\vecB{v}$ is the advection of the velocity from the magnetic field fluctuations. 

\section{The compressible quasi-static fluctuations model in the limit of large $B_0$}\label{appendix:vGradient}
    For $B_0 \gg \delta B$ Equation~(\ref{eq:induction2}) simplifies significantly. The time derivative of $\delta B$, on the LHS, is approximately zero, and so are the terms that have strict $\delta B$ dependence on the RHS. Hence
    \begin{align}
        0 &= B_0 \partial_{\parallel} \vecB{v} - \vecB{B}_0 (\nabla \cdot \vecB{v}), \\
        \partial_{\parallel} \vecB{v} &= \frac{\vecB{B}_0}{B_0} (\nabla \cdot \vecB{v}). 
    \end{align}
    The $\vecB{B}_0/B_0$ term is just the unit vector of $\vecB{B}_0$, $\hat{\vecB{B}_0}$, which contributes nothing to the magnitude, but sets direction of the compressive modes. Hence the magnitude of the compression is equal to the magnitude of the velocity gradient along the magnetic field,
    \begin{align}
        |\nabla \cdot \vecB{v} | = | \partial_{\parallel} \vecB{v} |,
    \end{align}
    as shown in Equation~(\ref{eq:induction2.1}), in the main text.
    
    \section{Weak $\M$ dependence of mean-field weighted fluctuations}\label{appendix:meanFieldWeighted}
\begin{figure*}
        \centering
        \includegraphics[width=\linewidth]{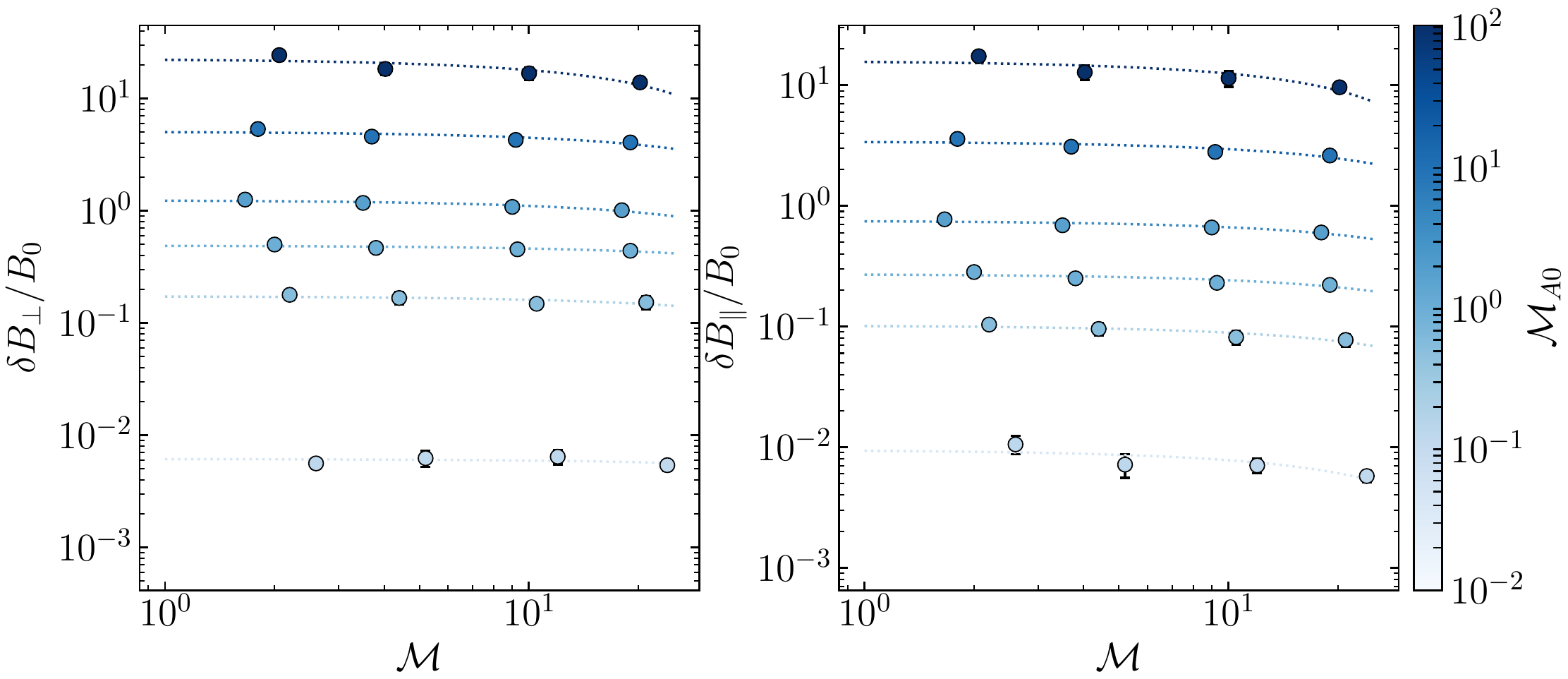}
        \caption{The same as Figure \ref{fig:perpAndpar}, but for the mean-field weighted fluctuations as discussed in \S\ref{sec:meanFieldWeighted}.}
        \label{fig:machDepend_meanfield}
    \end{figure*}

    Figure \ref{fig:machDepend_meanfield} shows how the perpendicular (left) and parallel (right) magnetic field fluctuations only very weakly depend upon $\M$. We discuss this result in \S\ref{sec:meanFieldWeighted}.
    
\begin{figure}
        \centering
        \includegraphics[width=\linewidth]{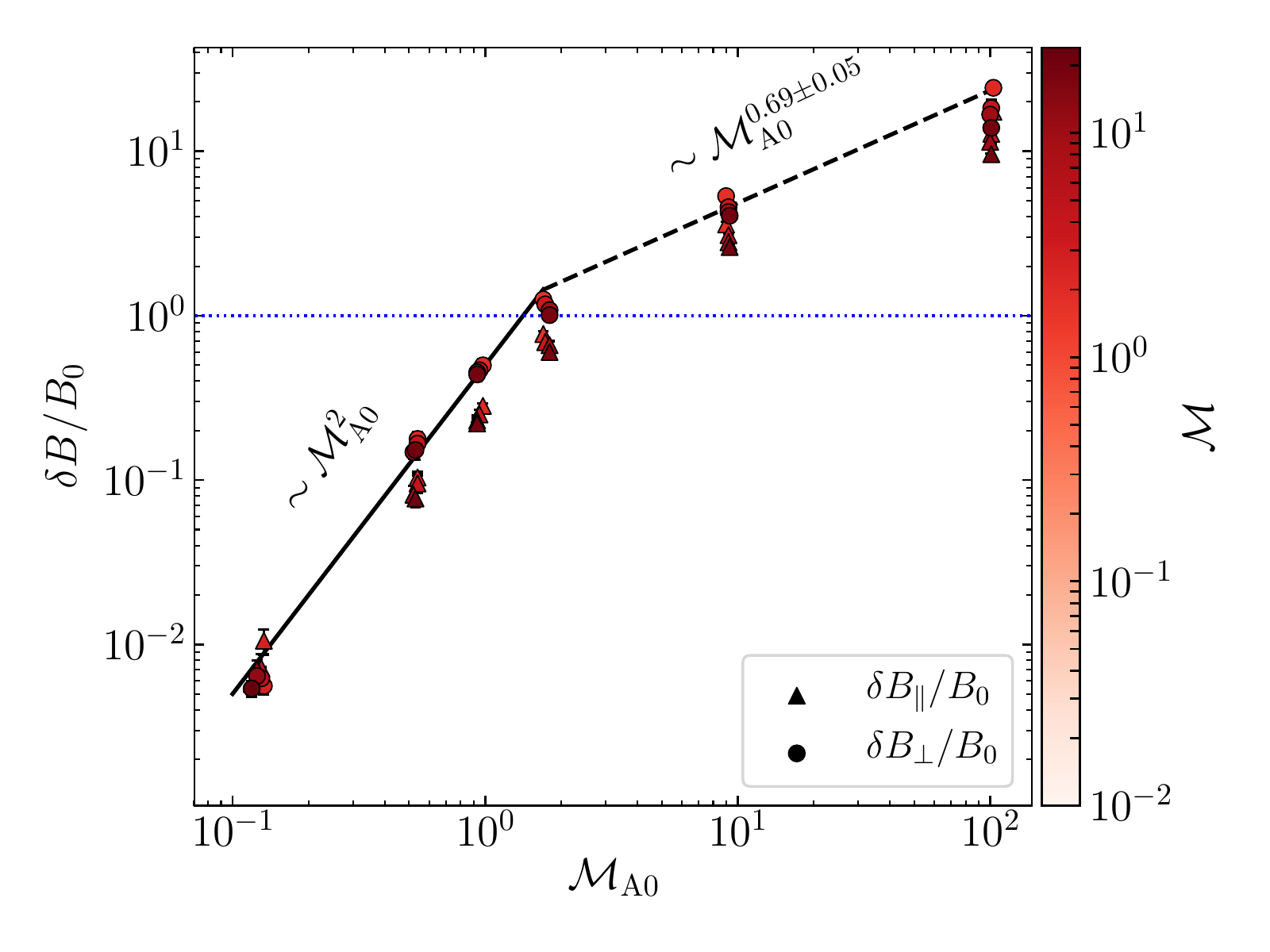}
        \caption{The same as Figure \ref{fig:weightedFluc}, but for $\delta B_{\parallel} / B_0$ (triangles) and $\delta B_{\perp} / B_0$ (circles). We see both of the magnetic field components follow a similar trend as $|\delta\vecB{B}| / B_0$, as discussed in \S\ref{sec:meanFieldWeighted}. }
        \label{fig:componentsWeighted}
    \end{figure}

    Figure \ref{fig:componentsWeighted} shows the $\delta B_{\parallel} / B_0$ (triangles) and $\delta B_{\perp} / B_0$ fluctuations. We show that they follow a similar trend as the magnitude of the full 3D fluctuations, $|\delta\vecB{B}| / B_0$, as discussed in \S\ref{sec:meanFieldWeighted}.
    
\bsp	
\label{lastpage}
\end{document}